\documentclass[11pt]{article}

\usepackage{amsfonts,amssymb,amsmath,latexsym,ae,aecompl}

\usepackage{graphicx}

\newcommand{\FF}{\vspace*{\medskipamount}}

\newcommand{\BBB}{\vspace*{-\bigskipamount}}

\newcommand{\cO}{\mathcal{O}}

\newcommand{\qed}{\hfill $\square$}
\newcommand{\Paragraph}[1]{\BBB\paragraph{#1}}

\newcommand{\RB}{\raisebox{3ex}{~}}
\newcommand{\LB}{\raisebox{-1.5ex}{~}}

\newtheorem{conjecture}{Conjecture}
\newtheorem{theorem}{Theorem}
\newtheorem{lemma}{Lemma}

\newenvironment{proof}{\noindent\textbf{Proof: }}{\qed \smallbreak}

\newlength{\pagewidth}
\newlength{\captionwidth}
\begin{document}

\baselineskip 	3ex
\parskip 		1ex

\title{		Packet Latency of Deterministic Broadcasting \\  
		in Adversarial Multiple Access Channels~\footnotemark[1]
		\vfill}
	
\author{	Lakshmi Anantharamu~\footnotemark[2]
		\and
		Bogdan S. Chlebus~\footnotemark[2] \ \
		\and
		Dariusz R. Kowalski~\footnotemark[3] \ \
		\and
		Mariusz A. Rokicki~\footnotemark[3]}

\footnotetext[1]{The results of this paper appeared in a preliminary form in~\cite{AnantharamuCKR-INFOCOM10} and~\cite{AnantharamuCKR-SIROCCO11}.
}

\footnotetext[2]{Department of Computer Science and Engineering,
University of Colorado Denver, Denver, Colorado 80217, USA.
The work of this author was supported by the National Science Foundation under Grant No. 1016847.}

\footnotetext[3]{Department of Computer Science,
University of Liverpool, Liverpool L69~3BX, United Kingdom.}

\date{}

\maketitle

\vfill

\begin{abstract}
We study broadcasting in multiple access channels with dynamic packet arrivals and jamming.
Communication environments are represented by adversarial models that specify constraints on packet arrivals and jamming. 
We consider deterministic distributed broadcast  algorithms and give upper bounds on the worst-case packet latency and the number of queued packets in relation to the parameters defining adversaries. 
Packet arrivals are determined by a rate of injections and a number of packets that can be generated  in one round. 
Jamming is constrained by a rate with which an adversary can jam rounds and by a number of consecutive rounds that can be jammed. 

\vfill

\end{abstract}

\vfill

\thispagestyle{empty}

\setcounter{page}{0}

\newpage

\section{Introduction}
\label{sec:introduction}

We study broadcasting in multiple access channels by deterministic distributed algorithms.
The communication medium may experience a mild form of jamming.
We evaluate the performance of communication algorithms by upper bounds on their packet latency (delay) and the number of packets queued at stations (queues size).
The performance metrics are understood in their worst-case sense and are considered in  adversarial frameworks of  packet injection and jamming.
There are no statistical components in either algorithms or traffic generation. 

The traditional approach to distributed broadcasting in multiple access channels uses randomization to arbitrate for access to a shared medium.
Typical examples of randomized broadcast algorithms include backoff ones, like the binary exponential backoff employed in the Ethernet.
The enduring effectiveness of the Ethernet, as a real-world implementation of local area networks \cite{MetcalfeB76}, is a compelling evidence that randomized broadcasting can perform well in practice.

Using randomization in algorithms, intended as practical solutions to broadcasting, may appear to be inevitable in order to cope with bursty traffic.
Among the main challenges that broadcasting on a shared channel faces is resolving conflicts for access to the communication medium.
In real-world applications, most stations stay idle for most of the time, so that periods of inactivity are interspersed with unexpected bursts of activity by groups of stations configured unpredictably.
Randomness appears to be a most natural way to break symmetry in attempts to access a channel.
Since traffic demands are typically assumed to be unpredictable, the methodological underpinnings of key performance metrics of broadcasting, like queue sizes and packet delay, have traditionally been studied with stochastic assumptions in mind.
In a matching manner, simulations have been geared towards models of packet generation defined by stochastic constraints.
All these factors have historically contributed to a popular perception that randomness and stochastic assumptions are inevitable aspects of broadcasting in multiple access channels. 

This paper addresses the efficiency of deterministic broadcast algorithms for dynamic traffic demands.
Performance of algorithms is measured by packet delay and the number of queued packets pending transmission, while packet injection is constrained by formal adversarial models.
Studying algorithmic paradigms useful for deterministic distributed broadcasting, for dynamic  packet injection, is a topic interesting in its own sake. 
We do this in a model of continuous packet injection without any stochastic assumptions about how packets are generated and where and when they are injected.
This model, known as adversarial queueing, is an alternative to representing packet generation by stochastic constraints.
Adversarial queuing has proved useful in providing frameworks to study dynamic communication  while imposing only minimal constraints on traffic generation.
It is an important benefit of adversarial queuing to provide a methodology to assess the performance of deterministic algorithms by worst-case bounds, with respect to suitable metrics. 

Jamming in wireless networks can be understood as either malicious disruptions of communication medium or inadvertent effects occurring on the physical layer.
The former is an effect of foreign messages sent deliberately to hinder the flow of information by creating interferences of legitimate signals with such external disrupting transmissions.
An example of jamming in this sense is a degradation-of-service attack that produces dummy packets that interfere with legitimate packets.
The latter interpretation of jamming is about the physical layer affected by external factors, such as the supply of energy, weather, or crowded bandwidth.
A closely related motivation is to interpret jamming as inadvertent collision of signals with concurrent foreign communication.
This occurs when groups of stations pursue their independent communication tasks, and so for each group an interference caused by foreign transmissions is logically equivalent to jamming.
To make our picture simple, jamming is understood in this paper as purely logical, in that this is a symptom we have to take into account without deliberating its causes. 
There are no assumptions made to justify why a transmitted message is not heard on the channel, including any references to the physical layer, while a message should be heard since only one station transmits in the round. 
A jammed round has the same effect as one with multiple simultaneous transmissions of stations attached to a channel, in that stations cannot distinguish a jammed round from a round with multiple transmissions.

\Paragraph{A summary of the methodology and results.}

We investigate deterministic broadcast algorithms for dynamic packet injection.
No randomization is used in algorithms nor there is any stochastic component that affects packet injection in the considered communication environments.
The studied communication algorithms are distributed in that they are executed with no centralized control. 
The two performance metrics are the queues size (maximum total number of packets simultaneously stored in the queues at stations while pending transmission) and packet latency (maximum number of rounds spent by a packet in a queue from injection until a successful transmission).

A set of stations attached to a channel is fixed and their number $n$ is known, in that it can be used in codes of algorithms.
Stations are equipped with private queues, in which they can store packets until they are transmitted successfully.

We use the slotted model of synchrony, in which an execution of a communication algorithm is partitioned into rounds, so that a transmission of a message with one packet  takes one round.
All the stations attached to the channel are  activated in the same initial round, each with an empty queue. 

It is the assumed synchrony that allows to define the rate of injecting packets and  the rate of jamming rounds.
A round comprises a short atomic duration of time during which some events happening in the system can be considered as occurring simultaneously.
For example, the burstiness of traffic is understood as the maximum number of packets that can be injected simultaneously, meaning in one round.
The related concept of burstiness of jamming is understood as the maximum number of contiguous rounds that are unavailable for successful transmissions because of continuous jamming. 
Similarly, it takes a full round to transmit a message.

We consider broadcasting against adversaries that control both injections of packets into stations and jamming of the communication medium.
Packet injection is  limited only by the rate of injecting new packets and the number of packets that can be injected simultaneously.
Jamming is limited by the rate of jamming different rounds and by how many consecutive rounds can be jammed.

All the considered algorithms have bounded packet latency for each fixed injection rate $\rho$ and jamming rate $\lambda$ subject only to the necessary constraint that $\rho+\lambda<1$.
The obtained upper bounds on packet latency and queue sizes of broadcast algorithms are understood in the worst-case sense.
Here  ``queue size'' means the maximum number of packets stored in the queues at the same time, as a function of $\rho$, for a given number~$n$ of stations, and packet latency is the maximum possible number of rounds spent by a packet in a queue waiting to be heard on the channel.

The upper bounds on queue size and packet latency of the algorithms studied in this paper are summarized in Tables~\ref{tab:no-jamming} and~\ref{tab:jamming}.
All the algorithms we consider are reviewed in detail in Section~\ref{sec:review-of-deterministic-broadcast-algorithms}.


\begin{table}[t]
\begin{center}
\begin{tabular}{|l |c |c |c|c|}
\hline
\RB \LB
Algorithm &Queues &Latency& Injection& Proved\\
\hline
\hline
\RB \LB
\textsc{OF-RRW}  & $\frac{2\rho }{1-\rho}\cdot n +\beta$&$\frac{2}{1-\rho}\cdot n +\beta(1+\rho)$ &$\rho<1$ & Thm~\ref{thm:OF-RRW} Sec~\ref{sec:non-adaptive-algorithms-without-jamming}\\
\hline
\RB \LB
\textsc{RRW} \cite{ChlebusKR-TALG12}&$\frac{2\rho }{1-\rho}\cdot n +\beta$ & $\frac{2-\rho}{(1-\rho)^2} \cdot n+ \frac{\beta}{1-\rho}$ &$\rho<1$& Thm~\ref{thm:RRW} Sec~\ref{sec:non-adaptive-algorithms-without-jamming}\\
\hline
\RB \LB
\textsc{OF-SRR} & $\frac{4\rho }{1-\rho}\cdot n  +\beta$  & $\frac{4}{1-\rho}\cdot n +\beta(1+\rho)$  & $\rho<1$& Thm~\ref{thm:OF-SRR} Sec~\ref{sec:non-adaptive-algorithms-without-jamming}\\
\hline
\RB \LB
\textsc{OF-SRR} & $2\beta$& $2\beta(2+\lg n)$ & $\rho \le \frac{1}{2+\lg n}$& Thm~\ref{thm:OF-SRR} Sec~\ref{sec:non-adaptive-algorithms-without-jamming}\\
\hline
\RB \LB
\textsc{SRR} \cite{ChlebusKR-TALG12} & $\frac{4\rho }{1-\rho}\cdot n  +\beta$& $\frac{4-2\rho}{(1-\rho)^2}\cdot n +\frac{\beta}{1-\rho} $ &$\rho<1$& Thm~\ref{thm:SRR} Sec~\ref{sec:non-adaptive-algorithms-without-jamming}\\
\hline
\RB \LB
\textsc{SRR} \cite{ChlebusKR-TALG12} & $2\beta$& $3\beta(2+\lg n)$ &$\rho \le \frac{1}{2+\lg n}$& Thm~\ref{thm:SRR} Sec~\ref{sec:non-adaptive-algorithms-without-jamming}\\
\hline
\RB \LB
 \textsc{MBTF} \cite{ChlebusKR09} & $\rho (1+\rho)\cdot n^2 + \beta $ & $\frac{1+\rho-\rho^2}{1-\rho} \cdot n^2 + \frac{\beta}{1-\rho}$ &$\rho<1$& Thm~\ref{thm:MBTF} Sec~\ref{sec:adaptive-algorithms-without-jamming}\\
\hline
\end{tabular}
\parbox{\pagewidth}{\FF\caption{\label{tab:no-jamming} 
Upper bounds on queue size and packet latency for a channel   without jamming with $n$ stations, executed against an adversary of injection rate $\rho<1$ and burstiness~$\beta\ge 1$. 
Algorithm  \textsc{MBTF}  is adaptive, and the remaining four algorithms are non-adaptive.
}}
\end{center}
\end{table}

We consider non-adaptive algorithms for channels without jamming when either collision detection is not available  (algorithms \textsc{OF-RRW} and \textsc{RRW}) or when it is available (algorithms \textsc{OF-SRR} and \textsc{SRR}).
These algorithms have a property that queue sizes grow unbounded with injection rate~$\rho$ approaching~$1$, for a fixed $n$.
We conjecture that this is a general phenomenon.


\begin{conjecture}
Each non-adaptive algorithm for channels without jamming that provides bounded queues, for injection rate~$\rho<1$, has its queue bound grow arbitrarily large as a function of injection rate~$\rho$, if $\rho$  approaches~$1$, for all sufficiently large and fixed numbers of stations~$n$.
\end{conjecture}

Adaptive algorithm \textsc{MBTF} for channels without jamming has bounded queues even when $\rho=1$, but its packet latency grows unbounded when $\rho<1$ approaches $1$.
We conjecture that this reflects a general property of broadcast algorithms.


\begin{conjecture}
Each broadcast algorithm for channels without jamming that provides bounded packet latency, for injection rate~$\rho<1$, has its packet-latency bound grow arbitrarily large as a function of injection rate~$\rho$, if $\rho$  approaches~$1$, for all sufficiently large and fixed numbers of stations~$n$.
\end{conjecture}

We show that a non-adaptive algorithm for channels with jamming achieves bounded packet latency for $\rho+\lambda<1$ when an upper bound on jamming burstiness is a part of code.
We hypothesize that this is unavoidable and reflects the utmost power of non-adaptive algorithms.


\begin{conjecture}
\label{hyp:fulls-sensing}

Each non-adaptive broadcast algorithm for channels with jamming  can be made unstable by some  adversaries with injection rates~$\rho$ and jamming rates~$\lambda$ satisfying $\rho + \lambda < 1$, for all sufficiently large  and fixed numbers of stations~$n$.
\end{conjecture}

Adaptive algorithm \textsc{C-MBTF} for channels with jamming has bounded queues when $\rho+\lambda=1$ but its packet latency increases unbounded when $\rho+\lambda<1$ approaches~$1$, for a fixed~$n$; see the discussion following the proof of Theorem~\ref{thm:CMBTF-jamming} in Section~\ref{sec:adaptive-algorithms-with-jamming} for details.
We conjecture that this is a general phenomenon.


\begin{conjecture}
Each broadcast algorithm for channels with jamming that provides bounded packet latency, for injection rate $\rho$ and jamming rate $\lambda$ such that $\rho+\lambda<1$, has its packet-latency bound grow arbitrarily large as a function of injection rate~$\rho$ and jamming rate $\lambda$, if $\rho+\lambda$  approaches~$1$, for all sufficiently large and fixed numbers of stations~$n$.
\end{conjecture}


\begin{table}[t]
\begin{center}
\begin{tabular}{|l |c |c |c|}
\hline
\RB \LB
Algorithm & Queues & Latency & Proved\\
\hline
\hline
\RB \LB
\textsc{OF-JRRW($J$)}& $\frac{2(\beta+1)}{1-\rho-\lambda} \cdot n+\beta$& $\frac{2(\beta+1)}{(1-\lambda)(1-\rho-\lambda)}\cdot n + \frac{\beta(1+\rho-\lambda)}{(1-\lambda)^2}$& Thm~\ref{thm:OF-JRRW(J)-jamming} Sec~\ref{sec:non-adaptive-algorithms-with-jamming}\\
\hline
\RB \LB
\textsc{JRRW}($J$) & $\frac{2(\beta+1)}{1-\rho-\lambda}\cdot n +\beta$& $\frac{2(\beta+1)}{(1-\lambda)(1-\rho-\lambda)^2}\cdot n + \frac{\beta(1-\lambda)}{1-\rho-\lambda}$ &Thm~\ref{thm:JRRW(J)-jamming} Sec~\ref{sec:non-adaptive-algorithms-with-jamming}\\
\hline
\RB \LB
\textsc{OFC-RRW} & $\frac{2\rho}{1-\rho-\lambda}\cdot n +\beta$ &$\frac{2}{1-\rho-\lambda} \cdot n+ \frac{\beta(1+\rho-\lambda) }{(1-\lambda)^2}$&Thm~\ref{thm:OFC-RRW-jamming} Sec~\ref{sec:adaptive-algorithms-with-jamming}\\
\hline
\RB \LB
 \textsc{C-RRW} &$\frac{2\rho}{1-\rho-\lambda}\cdot n +\beta$ &$\frac{2(1-\lambda)}{(1-\rho-\lambda)^2}\cdot n + \frac{\beta(1-\lambda)}{1-\rho-\lambda}$ &Thm~\ref{thm:C-RRW-jamming} Sec~\ref{sec:adaptive-algorithms-with-jamming}\\
\hline
\RB \LB
 \textsc{C-MBTF} &$\frac{\rho(1-\lambda)+\rho^2 }{(1-\lambda)^2}\cdot n^2 + \beta$ &$\frac{1+\rho-\lambda -\rho^2-2\rho\lambda}{(1-\lambda)(1-\rho-\lambda)}  \cdot n^2 + \frac{\beta(1-\lambda)}{1-\rho-\lambda}$ & Thm~\ref{thm:CMBTF-jamming} Sec~\ref{sec:adaptive-algorithms-with-jamming}\\
\hline
\end{tabular}
\parbox{\pagewidth}{\FF\caption{\label{tab:jamming} 
Upper bounds on queue size and packet latency for a channel with jamming  with $n$ stations,  when the injection and jamming rates satisfy $\rho+\lambda<1$ and for burstiness~$\beta\ge 1$.
The jamming burstiness is assumed to be at most~$J$ for algorithms \textsc{OF-JRRW($J$)} and \textsc{JRRW($J$)}, where $J$ is part of their codes.
Algorithms \textsc{OF-JRRW($J$)} and \textsc{JRRW($J$)} are non-adaptive, and the remaining three algorithms are adaptive.
}}
\end{center}
\end{table}


\Paragraph{Previous work on adversarial multiple access channels.}

Now we review previous work on broadcasting in multiple-access channels in the framework of adversarial queuing.
The first such work, by Bender et al.~\cite{BenderFHKL05}, concerned the throughput of randomized backoff for multiple-access channels, considered in the queue-free model.
Deterministic distributed broadcast algorithms for multiple-access channels, in the model of  stations with queues, were first considered  by Chlebus et al.~\cite{ChlebusKR-TALG12}; that paper 
 specified the classes of acknowledgment based and full sensing deterministic distributed algorithms, along the lines of the respective randomized protocols~\cite{Chlebus-randomized-radio-chapter-2001}.

The maximum throughput, defined to mean the maximum rate for which stability is achievable, was studied by Chlebus et al.~\cite{ChlebusKR09}.
Their model was of a fixed set of stations with queues, whose number~$n$ is known.
They developed a stable deterministic distributed broadcast algorithm with  queues of sizes that are  $\cO(n^2+\text{burstiness})$ against leaky-bucket adversaries of injection rate~$1$.
That work demonstrated that throughput~$1$ was achievable in the model of a fixed set of stations whose number $n$ is known.
The paper~\cite{ChlebusKR09} also showed some restrictions on traffic with throughput~$1$; in particular,  communication algorithms have to be adaptive (may use control bits in messages),  achieving bounded packet latency is impossible, and queues have to be of $\Omega(n^2+\text{burstiness})$ sizes.

Anantharamu et al. \cite{AnantharamuCR-TCS17} extended work on throughput~$1$ in adversarial settings by studying the impact of limiting window-type adversaries by assigning individual rates of injecting data for each station.
That paper~\cite{AnantharamuCR-TCS17} gave  a non-adaptive algorithm for channels without collision detection of $\cO(n+w)$ queue size and $\cO(nw)$ packet latency, where $w$ is the window size; this is in contrast with general adversaries, against whom bounded packet latency for injection rate~$1$ is impossible to achieve.

Bie{\'n}kowski et al.~\cite{BienkowskiJKK-DISC12} studied online broadcasting against  adversaries that are unbounded in the sense that they can inject packets into arbitrary stations with no constraints on their numbers nor rates of injection.
Paper~\cite{BienkowskiJKK-DISC12} gave a deterministic algorithm optimal with respect to competitive performance,  when measuring either the total number of packets in the system or the maximum queue size.
This algorithm was also shown in~\cite{BienkowskiJKK-DISC12} to be  stochastically optimal for any expected injection rate smaller than or equal to~$1$.

Anantharamu and Chlebus~\cite{AnantharamuC15} considered an ad-hoc multiple access channel, which has an unbounded supply of anonymous stations attached but only the stations activated  with injected packets participate in broadcasting.
They studied deterministic distributed broadcast algorithms against adversaries that are restricted to be able to activate at most one station per round.
The algorithms given in~\cite{AnantharamuC15} can provide bounded packet latency for injection rates up to $1/2$, with specific rates depending on additional features of algorithms.
It was also shown in~\cite{AnantharamuC15} that no injection rate greater than~$\frac{3}{4}$ can be handled with bounded packet latency on such ad-hoc channels by deterministic algorithms.


\Paragraph{Related work.}

A natural basic communication problem in multiple access channels concerns collision resolution:  there is a group of active stations, being a subset of all stations connected to the channel, and we want to have either some station in the group or all of them transmit successfully at least once.
For the recent work on this topic, see the papers by Kowalski~\cite{Kowalski05}, Fernandez Anta et al.~\cite{FernandezMM-Algorithmica13}, and De Marco and Kowalski~\cite{DeMarcoK15}.

Most related work on broadcasting in multiple access channels has been carried out with  randomization playing an integral part; see the survey~\cite{Chlebus-randomized-radio-chapter-2001}.
Randomness can affect the behavior of protocols either directly, by being a part of the mechanism of a communication algorithm, or indirectly, when packets are generated subject to stochastic constraints.
With randomness affecting communication in either way, the communication environment can be represented as a Markov chain with stability understood ultimately as ergodicity.
Stability of randomized communication algorithms can be considered in  the queue-free model, in which a  packet gets associated with a new station at the time of injection, and the station dies after the packet has been heard on the channel.
Full sensing protocols were shown to fare well in this model; some protocols stable for injection rate slightly below~$1/2$ were developed, see~\cite{Chlebus-randomized-radio-chapter-2001}.
The model of a fixed set of stations with private queues was considered to be less radical, as queues appear to have a stabilizing effect.
H\aa stad et al.~\cite{HastadLR96}, Al-Ammal et al.~\cite{Al-AmmalGM01} and  Goldberg et al.~\cite{GoldbergJKP04} studied bounds on the rates for which the binary exponential backoff was stable, as functions of the number of stations.
For recent work related to the exponential backoff see the papers by  Bender et al~\cite{BenderFGY16} and Bender et al~\cite{BenderKPY16}, who proposed modifications to exponential backoffs with the goal to improve some of their characteristics. 
Raghavan and Upfal~\cite{RaghavanU98} and Goldberg et al~\cite{GoldbergMPS-JACM00} proposed randomized broadcasts based on different paradigms that those used in backoff algorithms.

The methodology of adversarial queuing allows to capture the notion of stability of communication algorithms without resorting to randomness and can serve as a framework for worst-case bounds on performance of deterministic algorithms.
Borodin et al.~\cite{BorodinKRSW01} proposed this approach in the context of routing algorithms in store-and-forward networks. 
This was followed by Andrews et al.~\cite{AndrewsAFLLK01}, who emphasized the notion of universality in adversarial settings.

The adversarial approach to modeling communication proved to be inspirational and  versatile.
\'Alvarez et al.~\cite{AlvarezBDSF05} applied adversarial models to capture phenomena related to routing of packets with varying priorities and failures in networks.
\'Alvarez et al.~\cite{AlvarezBS-ICPADS04} addressed the impact of link failures on stability of communication algorithms by way of modeling them in adversarial terms.
Andrews and Zhang~\cite{AndrewsZ03} considered adversarial networks in which nodes operate as switches connecting inputs with outputs, so that routed packets encounter additional congestion constrains at nodes when they compete with other packets for input and output ports and need to be queued when delayed.
Andrews and Zhang~\cite{AndrewsZ07} investigated routing and scheduling in adversarial wireless networks in which every node can transmit data to at most one neighboring node per time-step and where data arrivals and transmission rates are governed by an adversary.

Worst-case packet latency of routing in store-and-forward wired networks has been studied in the framework of adversarial queuing.
Aiello et al.~\cite{AielloKOR-JCSS00} demonstrated that polynomial packet latency can be achieved by a distributed algorithm even when the adversaries do not disclose the paths they assigned to packets to validate complying with congestion constraints.
Andrews et al.~\cite{AndrewsFGZ-JACM05} studied packet latency of adversarial routing when the entire path of a packet is known at the source.
Broder et al.~\cite{BroderFU01} discussed conditions under which protocols effective for static routing provide bounded packet latency when applied in dynamic routing.
Scheideler and V\"ocking~\cite{ScheidelerV98} investigated how to transform static store-and-forward routing algorithms, designed to handle packets injected at the same time,  into efficient algorithms able to handle packets injected continuously into the network, so that packet delays in the static case are close to those occurring in the dynamic case.
Ros\'en and Tsirkin~\cite{RosenT06} studied bounded packet delays against the ultimately powerful adversaries of rate~$1$.

Jamming in multiple-access channels and wireless networks is usually understood as disruptions occurring in individual rounds that prevent successful transmissions in spite of lack of collisions caused by concurrent interfering transmissions.
Awerbuch~et al.~\cite{AwerbuchRSSZ14} studied jamming in multiple access channels in an adversarial setting with the goal to estimate saturation throughput of randomized protocols.
Richa et al.~\cite{RichaSSZ-TN13} gave a randomized medium-access algorithm against adaptive adversarial jamming of a shared medium that achieves a constant-competitive throughput. 
Gilbert~et al.~\cite{GilbertKPPSY14} studied jammed transmissions in multiple access channel with the goal to optimize energy consumption per each transmitting station.
Broadcasting in multi-channels with  jamming controlled by adversaries was studied by Chlebus et al.~\cite{ChlebusDK16}, Gilbert et al.~\cite{GilbertGKN-INFOCOM09}, Gilbert~et al.~\cite{GilbertGN09},  and Meier~et al.~\cite{MeierPSW09}.
Richa et al.~\cite{RichaSSZ-DC13} considered broadcasting on wireless networks modeled as unit disc graphs with one communication channel, in which a constant fraction of rounds can be jammed.

Jamming in multiple access channels is a special case of faulty behavior of wireless  networks.
Developing efficient fault-tolerant distributed communication algorithms in such networks has been an area of active investigations recently, of which the following is a sample.
 Alistarh et al.~\cite{AlistarhGGMN-SPAA10} studied non-cryptographic authenticated broadcast in radio networks when nodes are corrupted and behave in an unpredictable manner.
Bertier et al.~\cite{BertierKT-ICDCS10} designed message-efficient broadcast tolerating
Byzantine faults in a multi-hop wireless sensor networks.
Gilbert and Zheng~\cite{GilbertZ-TOPC15} proposed a protocol for downloading data from a single base station that is resilient to a sybil attack, during which multiple fake identities are simulated.
King et al.~\cite{KingSY-PODC11} studied communication channels that can be blocked by an adaptive adversary and proposed cost-efficient Las Vegas algorithms to send a message.
Ogierman et al.~\cite{OgiermanRSSZ14} 	considered wireless media under the SINR model subject to adversarial jamming of nodes and gave a randomized distributed medium-access  algorithm that achieves a constant competitive throughput.
Richa et al.~\cite{RichaSSZ-PODC12} studied multiple co-existing networks sharing a communication medium subject to adversarial jamming and gave a randomized medium-access algorithm to  effectively use the non-jammed rounds.
Tan et al.~\cite{TanWNS14} developed randomized solutions for multi-communication primitives in multi-hop multi-channel networks subject to adversarial disruptions of the shared channels. 
Young and Boutaba~\cite{YoungB11} surveyed the recent work on models and algorithms coping with faults in wireless communication, which includes adversarial jamming.

\Paragraph{Structure of the document.}

We review the model of multiple-access channels and summarize the classes of adversaries and deterministic broadcast algorithms in Section~\ref{sec:preliminaries}.
Section~\ref{sec:review-of-deterministic-broadcast-algorithms} contains a description of all the deterministic broadcast algorithms we consider, both old and new.
The analysis of performance of  broadcast algorithms is given in subsequent sections.
These are Section~\ref{sec:non-adaptive-algorithms-without-jamming} about non-adaptive algorithms for channels without jamming, Section~\ref{sec:adaptive-algorithms-without-jamming} about adaptive algorithms for channels without jamming, Section~\ref{sec:non-adaptive-algorithms-with-jamming} about non-adaptive algorithms for channels with jamming, and Section~\ref{sec:adaptive-algorithms-with-jamming} about adaptive algorithms for channels with jamming.
The final Section~\ref{sec:conclusion} includes a concluding discussion.

\section{Preliminaries}

\label{sec:preliminaries}

In this section, we review the model of multiple access channels and adversarial packet injection.
The considered communication environments allow to develop efficient deterministic distributed broadcast  algorithms.

A communication medium is called a \emph{channel}.
There are a number of communicating units attached to such a channel, which are called \emph{stations}.

We use the slotted model of synchrony, in which time is partitioned into \emph{rounds}.
The stations have access to a global clock measuring rounds, starting from round zero.
An execution of a communication algorithm starts with all the stations activated in this round zero.

The stations receive packets continuously and their goal is to have each of them eventually broadcast.
Each station is equipped with a private buffer space to store packets pending transmission.
Such a buffer is considered to have unbounded capacity, in that it can accommodate an arbitrary finite number of packets. 
The buffer memory of a station typically operates under a fixed queuing discipline and is referred to as a \emph{queue} of this station. 

A \emph{message} transmitted by a station on the channel may include at most one packet and it may include auxiliary control bits to coordinate actions of the stations.
The size of messages and the duration of rounds are calibrated such that a transmission of a message takes one round; this means that a station can transmit at most one message in a round.
Two messages transmitted by different stations in the same round overlap in time and are said to be \emph{transmitted simultaneously}.

A successful transmission of a message on the channel means that the message gets broadcast to all the stations.
If a message is delivered to a station then we say that that the message is \emph{heard} by the station.
If a message is heard by one station then it is also heard by all the stations. 
A round when no message is heard on the channel is called \emph{void}.

A round may be \emph{jammed}, which disrupts the communication functionality of the channel in this round; a round that is not jammed is called \emph{clear}.
A jammed round is always void but a clear round merely makes it possible to hear a message on the channel.

A communication environment we consider operates as a broadcast network consisting of ``active'' stations, which execute communication algorithms in a distributed manner, and a ``passive'' channel available for each station.
The ``external world'' uses such a communication environment by providing packets, which are injected individually into the stations, and it also determines which round is jammed.


\Paragraph{Multiple access channels.}

Broadcast networks we consider allow for jamming in general, but we also consider the case when no round can be jammed.
A broadcast network is said to be a \emph{multiple-access channel without jamming} when no round is ever jammed and a message transmitted by a station is heard if and only if it is the only message transmitted in the round. 
A broadcast network is said to be a \emph{multiple-access channel with jamming} when some rounds may be jammed and a message transmitted by a station is heard if and only if it is the only message transmitted in the round and the round is not jammed. 

In every round, all the stations receive feedback from the channel.
The feedback in a round is the same for each station; in particular, we do not differentiate between stations that transmit in a round and those that do not. 
If a message is heard on the channel, then the message itself is such a feedback.
A round with no transmissions is said to be \emph{silent}; in such a round,  all the stations receive from the channel the feedback we call \emph{silence}. 
Multiple transmissions in the same round result in conflict for access to the channel, which is called a \emph{collision}.
If a round is jammed then all the stations receive  in this round  the same feedback from the channel as in a round of collision.

Now we recapitulate all the possible reasons a round is void, that is, no message is heard.
One possibility is that the round is silent, in that there is no transmission.
The round may be jammed, then it does not matter whether there is any transmission in the round or not.
Finally, there may be a collision caused by multiple simultaneous transmissions.
Stations cannot distinguish between a round of collision, caused by multiple simultaneous transmissions, from a round in which the channel is jammed, in that the channel is sensed in exactly the same manner in both cases.

We say that \emph{collision detection} is available when stations can distinguish between silence and collision/jamming in a round by the feedback they receive from the channel in the round. 
If such a discerning mechanism is not available then the channel is \emph{without collision detection}.
Next we specify the four possible kinds of channels, determined by jamming or lack thereof, and, independently, by collision detection or lack thereof, which determine how stations perceive rounds by the obtained feedback from the channel.
\begin{description}
\item[\rm A channel without jamming and without collision detection:] 
a void round is caused by either silence or collision; a specific cause of voidness of a round is not perceivable.
\item[\rm A channel without jamming and with collision detection:] 
a void round is caused by either silence or collision; a specific cause of voidness of a round is identifiable.
\item[\rm A channel with jamming and without collision detection:] 
a void round is caused by either silence or collision or jamming; a specific cause of voidness of a round is not perceivable nor any can be excluded.
\item[\rm A channel with jamming and with collision detection:] 
a void round is caused by either silence or collision or jamming; silence can be perceived distinctly from the other two possible causes of voidness, but collision and jamming cannot be distinguished from each other.
\end{description}

A communication algorithm for channels without jamming can be executed on channels with jamming, without any changes in its code.
This is because a channel with jamming does not produce any special ``interference'' signal indicating that a round is jammed, and stations obtain either a silence or collision as  feedback from the channel when a round is void.


\Paragraph{An adversarial model of packet injection without jamming.}

We use a leaky-bucket adversarial model of packet injection, when a channel cannot be jammed, similarly as considered in \cite{AndrewsAFLLK01, ChlebusKR09}.
An adversary is determined by its maximum rate of injecting packets and a burstiness of traffic it can generate. 
Let a real number~$\rho$  and integer $\beta$ satisfy the inequalities $0<\rho\le 1$ and~$\beta\ge 1$; 
the \emph{leaky-bucket adversary of type $(\rho,\beta)$} may inject at most $\rho t+\beta$ packets into an arbitrary set of stations in each contiguous segment of $t>0$ rounds.
An adversary of type $(\rho,\beta)$ is said to have \emph{injection rate $\rho$} and \emph{burstiness component}~$\beta$.
The \emph{burstiness} of an adversary means the maximum number of packets that can be injected  in one round.
An adversary of type $(\rho,\beta)$ has burstiness~$\lfloor\rho+\beta\rfloor$, so if $\rho<1$ then $\beta$ is the adversary's burstiness.

In some broadcast algorithms, in which the place and time of injection of packets determines the order of their future transmissions, a prescribed quantity $k$ of rounds that occur allows the adversary to inject $\rho k$ packets, which then take $\rho k$ rounds to be transmitted, thus delaying transmissions of older packets.
If this pattern can be iterated, then this creates a combined delay of the following duration:
\[
k + \rho k + \rho^2 k+\cdots \le \frac{k}{1-\rho}
\ .
\]
We say that the quantity $\frac{k}{1-\rho}$ is obtained from $k$ by \emph{stretching-by-injecting}.


\Paragraph{An adversarial model of packet injection and jamming.}

For channels with jamming, we consider adversaries that control both packet injections and jamming.
Given real numbers $\rho$ and~$\lambda$ in the interval $(0,1]$ and integer~$\beta\ge 1$, the \emph{leaky-bucket jamming adversary of type $(\rho,\lambda,\beta)$} can inject at most $\rho t + \beta$ packets and, independently, it can jam at most $\lambda t + \beta$ rounds, in each contiguous segment of $t>0$ rounds.
For such an adversary, we refer to $\rho$ as the \emph{injection rate}, to $\lambda$ as the \emph{jamming rate}, and to $\beta$ as the \emph{burstiness component}. 
We can observe that a non-jamming adversary of type $(\rho,\beta)$ is formally the same as a  jamming adversary of type~$(\rho,0,\beta)$.
The number of packets that a jamming adversary can inject in one round is called its injection burstiness, similarly as for a non-jamming leaky-bucket adversary.
This parameter equals $\lfloor \rho + \beta\rfloor$.
If $\lambda= 1$ then every round could be jammed, making the channel dysfunctional.
Therefore, we always assume that a jamming rate~$\lambda$ satisfies $\lambda< 1$.

Suppose we are concerned about a contiguous segment of $k$ non-jammed rounds, possibly interspersed with $x$ additional jammed rounds.
If the adversary wants to stretch $k+x$ as much as possible by maximizing~$x$, then the inequality $\lambda (k+x) + \beta\ge x$ has to hold.
If this is applied repeatedly and the adversary jams at full power then the burstiness component~$\beta$ can be applied only once.
Disregarding the burstiness component~$\beta$ in the inequality $\lambda (k+x) + \beta\ge x$ is the same as setting $\beta=0$, so we have the inequality $\lambda (k+x)\ge x$, which gives $x\le \frac{\lambda }{1-\lambda}\cdot k$.
We obtain the following estimate 
\[
k+x\le k + k\cdot \frac{\lambda }{1-\lambda} = \frac{ k}{1-\lambda}
\ .
\]
We say that the quantity $\frac{ k}{1-\lambda}$ is obtained from $k$ by \emph{stretching-by-jamming}.

If the adversary injects with injection rate $\rho$ during these $k$ non-jammed rounds extended by inserted jammed rounds, then the number of injected packets in the whole interval that includes jammed rounds is at most the quantity 
\[
 \frac{ \rho}{1-\lambda} \cdot k
\ ,
\]
which is the same as if $\rho$ got expanded to a virtual injection rate $\frac{\rho}{1-\lambda}$ by an effect similar to stretching-by-jamming.
The quantity $\frac{\rho}{1-\lambda}$ can indeed be interpreted as injection rate because $\frac{\rho}{1-\lambda}<1$, as $\rho<1-\lambda$.
If the adversary applies this virtual injection rate, already obtained by stretching-by-jamming, by creating a stretching-by-inserting effect, an interval of $k$ clear rounds gets extended to the following number of rounds
\[
k\bigl(1 + \frac{\rho}{1-\lambda} +\bigl( \frac{\rho}{1-\lambda}\bigr)^2+\ldots\bigr) 
=  \frac{k}{1-\frac{\rho}{1-\lambda}} 
= \frac{k(1-\lambda)}{1-\rho-\lambda}
\ .
\]
We say that the quantity $\frac{k(1-\lambda)}{1-\rho-\lambda}$ is obtained from~$k$ by \emph{combined stretching}.

A maximum continuous number of rounds that an adversary can jam is called its \emph{jamming burstiness}.
We can find what is the jamming burstiness for a  leaky-bucket jamming adversary of type $(\rho,\lambda,\beta)$ as follows.
Let $x$ be a number of rounds that make a contiguous interval and are all jammed.
The inequality $\lambda x+\beta \ge x$ needs to hold, as otherwise $x$ rounds within an interval of~$x$ rounds could not be jammed. 
We conclude by algebra that the adversary can jam at most $\frac{\beta}{1-\lambda}$ consecutive rounds, which is an instance of stretching-by-jamming.


\Paragraph{Deterministic distributed broadcast algorithms.}

Broadcast algorithms control timings of transmissions by individual stations in a deterministic manner, starting from round zero when all the stations are activated simultaneously.
All the algorithms we consider are \emph{work-preserving} in that if a station is scheduled to transmit and it has pending packets then a transmitted message includes a packet.

A \emph{state} of a station is determined by the values of the private variables occurring in the code of an algorithm and by the number of outstanding packets in its queue that still need to be transmitted.
The local queues of packets at stations operate under the first-in-first-out discipline, which minimizes packet latency.
A  station obtains a packet to broadcast by removing the first packet from the queue.
If a station transmits a packet that is not heard then the station will transmit the same packet in the immediately following round in which a transmission is scheduled. 
A packet is never dropped by a station before it is heard on the channel.

A \emph{state transition} is a change in a state of a station in one round, which depends on the state at the end of the previous round, the feedback from the channel in this round, and the packets injected in this round.
A state transition of a station in a round  consists of the following  actions in order.
If packets are injected into the station in this round then they are immediately enqueued into the local queue.
If the station broadcasted successfully in the previous round, then the transmitted packet is discarded.
If a new packet to transmit is needed and the local queue is nonempty then a packet is obtained by dequeuing the queue.
Finally, a message for the next round is prepared, if any will be  transmitted.

An \emph{event} in a round comprises the following four actions by each  station in the given order: 
(a)~a~station either transmits a message or pauses, accordingly to its state, 
(b)~a~station receives a feedback from the channel, in the form of either hearing a message or collision signal or silence,  
(c)~new packets are injected into a station, if any, and finally,
(d)~the suitable state transition occurs at a station.
An \emph{execution} of an algorithm is a sequence of events occurring in consecutive rounds.

We categorize broadcast algorithms according to the terminology used in~\cite{ChlebusKR09, ChlebusKR-TALG12}.
All the algorithms considered in this paper are full sensing, in that nontrivial state transitions can occur at a station in any round, even when the station does not have pending packets to transmit.
This may be interpreted as if the attached stations  ``sense the channel'' in all rounds.
Algorithms that use control bits piggybacked on packets or can send messages comprised of only control bits, when a station does not have a packet to transmit, are called \emph{adaptive}, and otherwise they are \emph{non-adaptive}.


\Paragraph{Performance of broadcast algorithms.}

The basic quality for a communication algorithm in a given adversarial environment is \emph{stability},  understood to mean that the number of packets in the queues at stations stays uniformly bounded at all times.
For a stable algorithm in a communication environment, an upper bound on the number of packets waiting in queues is a natural performance metric, see~\cite{ChlebusKR09, ChlebusKR-TALG12}.

We may observe that stability is not achievable by a jamming adversary with injection rate~$\rho$ and a jamming rate~$\lambda$ satisfying $\rho+\lambda>1$. 
To see this, observe that it is equivalent to $\rho>1-\lambda$, so when the adversary is jamming with the maximum power, then the bandwidth remaining for transmissions is $1-\lambda$, while the injection rate is greater than~$1-\lambda$.

A sharper performance metric is that of \emph{packet latency}; it denotes an upper bound on the time spent by a packet waiting in a queue, counting from the round of injection through the round when the packet is heard on the channel. 
It is possible to achieve stability in the case $\rho+\lambda =1$, by adapting the approach for $\rho=1$ (and $\lambda=0$) in~\cite{ChlebusKR09}, but packet latency is then inherently unbounded.

An algorithm for an environment without jamming is \emph{universal} when it is stable for any injection rate smaller than~$1$.
This can be extended to jamming by having stability for each case of $\rho+\lambda < 1$.
All the algorithms we present are universal in this sense.
For each algorithm discussed in this paper, we give upper bounds for packet latency as functions of the number of stations~$n$ and the type $(\rho,\lambda,\beta)$ of a leaky-bucket (jamming) adversary, subject only to the restriction $\rho+\lambda < 1$.


\Paragraph{Knowledge.}

A property of a system is said to be \emph{known} when it can be referred to explicitly in codes of algorithms.
We assume throughout that the number of stations~$n$ is known to the stations.
Each station has a unique integer name in $[0,n-1]$, which it knows.
If a station needs to be distinguished in a communication algorithm, for example to be the first one to transmit in an execution, then by default it is the station with name~$0$.

The type of an adversary is normally not assumed to be known by the algorithms in this paper.
The only exception to this rule  occurs for a non-adaptive algorithm given in Section~\ref{sec:non-adaptive-algorithms-with-jamming} that has  an upper bound~$J$ on the jamming burstiness of an adversary as part of its code; this algorithm attains the claimed packet latency when the adversary's jamming  burstiness happens to be at most~$J$.

\section{A Review of Deterministic Broadcast Algorithms}

\label{sec:review-of-deterministic-broadcast-algorithms}

We summarize the specifications of deterministic distributed broadcast algorithms whose packet latency is analyzed in the following Sections.


\Paragraph{Three broadcast algorithms.}

We start with a summary of three deterministic distributed algorithms for channels without jamming that are already known in the literature.
These are the algorithms \textsc{RRW}, \textsc{SRR} and \textsc{MBTF}, which can be described as follows.

Algorithm \textsc{Round-Robin-Withholding} (\textsc{RRW}) is a  non-adaptive algorithm for channels without collision detection.
It operates in a round-robin fashion, in that the stations gain access to the channel in the cyclic order of their names.
A station with the right to transmit is said to hold a conceptual \emph{token}.
Once a station receives the token then it withholds the channel to unload all the packets in its queue.
A silent round is a signal for the next station, in the cyclic order of names,  to take over the token.
Algorithm \textsc{RRW} was introduced in~\cite{ChlebusKR-TALG12} and showed to be universal, that is, stable for injection rates smaller than~$1$.

Algorithm \textsc{Search-Round-Robin} (\textsc{SRR}) is a non-adaptive algorithm for channels with collision detection.
Its execution proceeds as a systematic continuous search for the next station with packets to transmit, under the cyclic ordering of stations by their names.  
The search is interpreted as binary one and is implemented by using a virtual distributed stack.
If a station with pending packets is identified by the search, the search is suspended while the station withholds the channel to transmit all its packets.
After all the packets held by a station have been unloaded, a silent round follows, which triggers the search to be resumed.
A basic step in searching is to verify if there is a station with pending packets whose name is in a given interval of integers.
Such a step is accomplished by all the stations in the interval transmitting their packets. 
Every station receives the same feedback from the channel, whether it transmitted or not, so all the stations know if the interval is empty (silence), or it contains a single station (packet heard), or it contains multiple stations (collision).
A search for the next station is completed by a packet heard.
A silence indicates that no station in the tested segment has packets and the interval is discarded. 
A collision results in having the interval partitioned into two halves of equal sizes, with one part processed immediately next while the other one is pushed on a stack to wait.
If a processed interval becomes empty or it is verified by silence that there is no station with packets in it, then a new interval is obtained by popping the stack.
One instance of a full sweep through all the stations is called a \emph{phase}.
A phase starts with the interval $[0,n-1]$ representing all the stations placed on the stack, and it ends with the stack becoming empty.
Once a phase is completed, the next similar phase begins immediately.
Algorithm \textsc{SRR} was  introduced in~\cite{ChlebusKR-TALG12} and showed to be universal.

Algorithm \textsc{Move-Big-To-Front} (\textsc{MBTF}) is an adaptive algorithm that can be executed on  channels without collision detection.
Each station maintains a dynamic list of all the stations in its private memory.
Such a list is initialized in each station to have all the names of stations arranged in the increasing order:  $0,1,2\ldots, n-1$.
The lists are manipulated in the same way by all the stations so they are identical copies of each other.
The algorithm schedules exactly one station to transmit in a round, so that collisions never occur.
This is implemented by having a conceptual token travel through the stations, which is initially assigned to the first station in the list.
A station with the token broadcasts a packet, if it has any, otherwise the round is silent.
A station considers itself \emph{big} in a round when it has at least $n$ packets; such a station attaches a control bit to every packet it transmits to indicate this status.
A big station is moved to the front of the list and it takes the token with it.  
If a station that is not big transmits in a round, or when it pauses  due to a lack of packets while holding the token so the round is silent, then the token is passed in this round to the next station in the list ordered in a cyclic fashion.
Algorithm \textsc{MBTF} was introduced in~\cite{ChlebusKR09} and showed to be stable for injection rate~$1$.


\Paragraph{The ``old-go-first'' approach.}

We obtain new algorithms by modifying \textsc{RRW} and \textsc{SRR} so that packets are categorized into ``old'' and ``new.''
Intuitively, packets categorized as ``new'' become eligible for transmissions only after all the packets categorized as ``old'' have been heard.
Formally, an execution is structured as a sequence of conceptual phases, which are contiguous segments of rounds of dynamic length, and then the notions of old versus new packets are defined with respect to them.

A \emph{phase} is defined as a full cycle made by the conceptual token visiting the stations.
No additional communication is needed to mark a transition to a new phase as all the stations can detect this by monitoring the position of the virtual token.
A token leaves a station holding it after the station has transmitted all its old packets while new packets may remain waiting for the next token's visit. 
In a given phase, packets are \emph{old} when they had been injected in the previous phase, and packets injected in the current phase are considered \emph{new} for the duration of the phase.
If a new phase begins, the old packets have already been heard on the channel and the new  ones immediately graduate to becoming old.
This means that the ``old-go-first'' principle is implemented by having packets injected in a given phase transmitted only in the next phase.
In particular, the first phase does not include any transmissions of packets, as all the packets, if any, are new.

Specifically, algorithm \textsc{Old-First-Round-Robin-Withholding} (\textsc{OF-RRW}) operates by  manipulating the token similarly as algorithm \textsc{RRW} does, except that when a station gets access to the channel by transmitting successfully, then the station unloads all the old packets, while new packets stay in the queue when the token is passed to the next station.
Algorithm \textsc{Old-First-Search-Round-Robin} (\textsc{OF-SRR}) performs search  similarly as algorithm \textsc{SRR} does, except that searching is for old packets only while new ones are ignored for the duration of a phase.
This approach is also applied to algorithm \textsc{JRRW$(J)$} for channels with jamming, as explained next.

The approach to modify a token algorithm by making old packets go first makes packet latency smaller than in the original version but queue bounds remain the same, as reflected by the bounds summarized in Tables~\ref{tab:no-jamming} and~\ref{tab:jamming}.
The difference in packet latency is such that a ``regular'' version of an algorithm for channel without jamming, which is either \textsc{RRW} or \textsc{SRR}, has an additional factor of~$\frac{1}{1-\rho}$ present in its bound on packet latency as compared to their versions with old-go-first specification, and the bound for algorithm \textsc{JRRW$(J)$} has an extra factor of $\frac{1}{1-\rho-\lambda }$ present, as compared to the bound on packet latency for algorithm \textsc{OF-JRRW$(J)$}.
This might be counter-intuitive, as an old-go-first version of broadcasting is a ``lazy'' implementation, in the sense that a possible immediate transmission of a packet is delayed for later when the packet happens to be still new.
This can be explained intuitively as follows.
Consider a regular version of a given broadcast algorithm, like \textsc{RRW}.
An injected packet may be transmitted either in the current phase or in the next phase, depending on how the station that the packet is injected into is located in the cycle of stations with respect to the station holding the token at the round of injection.
We may say that injecting ``behind the token'' results in transmitting in the next phase and injecting ``ahead of the token'' results in transmitting in the current phase. 
If the adversary consistently injects ``behind the token'' so that packets are transmitted as already old then a execution is indistinguishable from that of the old-go-first version of the algorithm.  
There is a possibility of an effect of stretching-by-injecting occurring in executions of the old-go-first version and this is reflected in the factor of~$\frac{1}{1-\rho}$ in the bound on packet latency.
If the adversary exercises the option to inject ``ahead-of-the-token,''  for the regular version of the algorithm, then this creates an additional possibility of enforcing  stretching-by-injecting, and so adds another factor of~$\frac{1}{1-\rho}$.


\Paragraph{Non-adaptive algorithms for channels with jamming.}

We introduce a non-adaptive broadcast algorithm \textsc{Jamming-Round-Robin-Withholding}$(J)$, abbreviated  \textsc{JRRW$(J)$}, for channels with jamming.
The design of the algorithm is similar to that of \textsc{RRW}, the difference is in how the  token is transferred from a station to the next one, in the cyclic order among the stations.
Just one void round should not trigger a transfer of the token, as it is the case in \textsc{RRW},  because not hearing a message may be caused by jamming.

The algorithm has a parameter~$J$ interpreted as an upper bound on jamming burstiness of the adversary.
This parameter is used to facilitate transfer of  control from a station to the next one  by way of forwarding the token.
The token is moved after precisely $J+1$ contiguous void rounds, counting from either hearing a packet or moving the token; the former indicates that the transmitting station exhausted its queue, while the latter indicates that the queue was empty.
More precisely, every station maintains a private counter of void rounds. 
The counters show the same value across the system, as they are updated in exactly the same way  determined only by the feedback from the channel.
A void round results in incrementing the counter by~$1$. 
The token is moved to the next station when the counter reaches~$J+1$.
If either a packet is heard or the token is moved then the counter is zeroed.

Algorithm \textsc{Old-First-Jamming-Round-Robin-Withholding}$(J)$, abbreviated \textsc{OF-JRRW$(J)$}, is obtained from \textsc{JRRW$(J)$} similarly as \textsc{OF-RRW} is obtained from~\textsc{RRW}.
An execution is structured as consisting of consecutive phases, and packets are categorized into old and new, with the same rule to graduate packets from new to old.
If a token visits a station, then only the old packets are transmitted while the new ones will be transmitted during the next visit by the token.


\Paragraph{Structural properties of algorithms.}

We say that a communication algorithm designed for a channel without jamming is a \emph{token} one if it uses a virtual token to determine a station that gains the right to transmit successfully.
All the algorithms discussed in this paper could be considered as token ones.
This is clearly the case for algorithms \textsc{RRW}, \textsc{OF-RRW}, \textsc{JRRW}, \textsc{OF-JRRW}, and \textsc{MBTF}, as  their design specifies how a token is handled.
Algorithms \textsc{SRR} and \textsc{OF-SRR} can also be interpreted as token ones, even though they make collisions possible to happen.
A station that transmits a packet successfully can be considered as holding the token, in that it can safely withhold the channel, and the right to transmit was acquired by the virtue of being the next station with packets after the previously transmitting one, in the cyclic ordering of stations.

A token algorithm for channels without collision detection and without jamming can be modified to the model with jamming, but still without collision detection.
This can be done in the following manner.
If a station has the right to transmit a packet in the original algorithm, then the modified algorithm has the station transmit a packet as well, otherwise the station transmits a control bit.
A round in which only a control bit is transmitted by a modified token algorithm is called a  \emph{control round} otherwise it is a \emph{packet round}.
The effect of sending control bits in control rounds is that if a round is not jammed then a message is heard in this round; this message is either just a control bit or it includes a packet.
This approach to replace silent rounds by rounds with messages with control bits allows for jamming detection: when a void round occurs  then this round has to be jammed, as otherwise a message would be heard.
Once a communication algorithm can identify jammed rounds, we may ignore their impact on the flow of control and repeat the performed actions in the next round, exactly as they were performed in the immediately preceding jammed ones. 
The resulting algorithm is clearly adaptive.
This method cannot be applied to algorithms relying on collision detection, like \textsc{SRR} and \textsc{OF-SRR}.

We will apply this method of modifying token algorithms to the non-adaptive algorithms \textsc{RRW} and \textsc{OF-RRW},  denoting the modified versions by \textsc{C-RRW} and \textsc{OFC-RRW}, respectively.
Similarly, we modify algorithm \textsc{MBTF} such that a station with a token sends a control message even if the station does not have a packet; the modified algorithm is denoted by \textsc{C-MBTF}.
The letter~C is a mnemonic to indicate using control rounds for jamming detection.

Algorithms with executions structured into phases, so that each station with packets has one opportunity to transmit its packets in a phase, are referred to as \emph{phase algorithms}. 
Among the algorithms considered in this paper, all are phase ones except for \textsc{MBTF} and \textsc{C-MBTF}.
The phase algorithms consist of \textsc{RRW}, \textsc{OF-RRW}, \textsc{C-RRW}, \textsc{OFC-RRW}, \textsc{SRR}, \textsc{OF-SRR}, \textsc{JRRW} and \textsc{OF-JRRW}.
If the old-go-first approach is used in a phase algorithm then it is  an  \emph{old-go-first version} of the algorithm, otherwise it is a \emph{regular version} of the algorithm.
In particular, \textsc{RRW},  \textsc{C-RRW}, \textsc{SRR} and \textsc{JRRW} are all regular phase algorithms, while \textsc{OF-RRW}, \textsc{OFC-RRW}, \textsc{OF-SRR} and \textsc{OF-JRRW} are all old-go-first phase algorithms.

Let us consider an execution of a  token algorithm.
If a packet is injected into a station whose number is smaller than that of the current token's holder then we say that the packet is injected \emph{behind the token}, and otherwise it is injected \emph{ahead of the token}.
If the considered token algorithm is a regular one, like \textsc{RRW}, then packets injected behind the token are transmitted in the next phase, and those injected ahead of the token are  transmitted in the current phase.

\section{Non-adaptive Algorithms without Jamming}

\label{sec:non-adaptive-algorithms-without-jamming}

In this Section, we consider deterministic distributed non-adaptive algorithms for channels without jamming for injection rates $\rho < 1$. 
For each of these algorithms, we give upper bounds for the queue size and packet latency as functions of the number of stations~$n$ and the type $(\rho,b)$ of a leaky-bucket adversary.

\subsection{Channels without collision detection}

We begin with algorithms \textsc{OF-RRW} and \textsc{RRW} for channels without collision detection. 
Each of them is a token algorithm.
The token is advanced to the next station when a station holding the token at the moment pauses, which results in a silent round.


\begin{theorem}
\label{thm:OF-RRW}

If algorithm \textsc{OF-RRW}  is executed  by $n$ stations  against an adversary of type~$(\rho,\beta)$  then the number of packets simultaneously queued in the stations is at most
\begin{equation}
\label{eqn:OF-RRW-queues}
\frac{2\rho }{1-\rho}\cdot n +\beta
\end{equation}
and packet latency is at most
\begin{equation}
\label{eqn:OF-RRW-latency}
\frac{2}{1-\rho}\cdot n +\beta(1+\rho)
\ .
\end{equation}
\end{theorem}

\begin{proof}
Let $T_i$ denote the duration of phase $i$, where $T_1=n$.
Let $Q_i$ denote the number of old packets in the beginning of phase~$i$, where $Q_1=0$.
The sequences $(Q_i)_{i\ge 1}$ and $(T_i)_{i\ge 1}$ satisfy the following recursive dependencies, where we disregard the effect of burstiness:
\[
Q_{i+1}\le \rho\cdot T_i
\]
and 
\[
T_{i+1}\le n+Q_{i+1}
\ ,
\]
by the algorithm's design and the constraints imposed on the adversary.
Iterating these recurrences produces the following bound $T$ on the duration of a phase:
\begin{equation}
\label{eqn:OF-RRW-no-jamming}
T_{i+1} \le n + \rho\cdot T_i \le n+ \rho n + \rho T_{i-1} \le n(1+\rho+\rho^2+\ldots)\le \frac{n}{1-\rho} =T
\ .
\end{equation}
A packet waits to be transmitted through at most two consecutive phases, each taking at most~$T$ rounds.
A bound for $T$ given in~\eqref{eqn:OF-RRW-no-jamming} disregards the effect of burstiness.
We can account for the effect of burstiness as follows.
Let the adversary inject additional $\beta$ packets in a round of a phase.
This instantaneously increases the number of packet queued in the current phase but extends the duration of the next phase, which is the phase when these packets are transmitted as old.
These transmissions in turn allow the adversary to inject $\rho \beta$ additional packets, which extends the duration of the next phase by $\rho \beta$ rounds.

We conclude with the following estimates.
The maximum number of queued packets is obtained by combining at most $\rho T$ old packets with  at most $\rho T$ new packets, along with at most $\beta$ packets injected in a burst, which together give~\eqref{eqn:OF-RRW-queues} as a bound.
The maximum number of rounds spent by a packet waiting to be heard on the channel is obtained by adding twice the upper bound~$T$ on a duration of a phase~\eqref{eqn:OF-RRW-no-jamming}, incremented by $\beta$ extra rounds in a phase immediately following one of a bursty injection, along with $\rho\beta$ rounds of the next phase, which together give~\eqref{eqn:OF-RRW-latency}.
\end{proof}

The bounds of Theorems~\ref{thm:OF-RRW}  are asymptotically tight. 
We give a strategy of the adversary to make queue sizes and packet latency close to these for algorithm \textsc{OF-RRW}.
When a phase begins then the adversary injects its first packet into station $n-1$, to make it wait almost two phases.
The adversary injects at full power, that is, as soon as a packet can be injected while satisfying the restriction that the number of packets injected is at most $\rho t$ within the first $t$ rounds of an execution, then a packet is injected.
The first phase takes exactly $n$ rounds, and the adversary injects $\rho n$ packets during this phase, but all of them will be transmitted in the next phase.
So when the second phase begins, there are already $\rho n$ packets queued.
The duration of phases keeps increasing such that when one takes $r$ rounds then the next one takes $r+\rho r$ rounds, starting from~$n$, so that it gets arbitrarily close to $\frac{n}{1-\rho}$.
The number of old packets is $\rho$ times the duration of a phase.
Burstiness allows to add $\beta$ to the number of queued packets and extend two consecutive phases by  $\beta(1+\rho)$ rounds. 

Next we estimate the performance of algorithm \textsc{RRW}.


\begin{theorem}
\label{thm:RRW}

If algorithm \textsc{RRW}  is executed by $n$ stations against an adversary of type $(\rho,\beta)$ then the number of packets simultaneously queued in the stations is at most
\begin{equation}
\label{eqn:RRW-queues}
\frac{2\rho }{1-\rho}\cdot n +\beta
\end{equation}
and packet latency is at most
\begin{equation}
\label{eqn:RRW-latency}
\frac{2-\rho}{(1-\rho)^2} \cdot n+ \frac{\beta}{1-\rho}
\ .
\end{equation}
\end{theorem}

\begin{proof}
First consider the queue sizes. 
Packets injected behind the token are transmitted in the next phase, which is consistent with the design of \textsc{OF-RRW} and so with its bound.
Packets injected ahead of the token are transmitted in the current phase, which slows down the phase compared to \textsc{OF-RRW}.
If a phase is longer then more packets can be injected in it, but each extra round is spent on a  transmission, because this is the reason a phase is longer, while not each extra round has to have a new packet injected in it.
This means that the upper bound on the number of packets stored in the queues~\eqref{eqn:OF-RRW-queues} derived for \textsc{OF-RRW} also applies to \textsc{RRW}, so we make it equal to~\eqref{eqn:RRW-queues}.

Next we estimate packet latency.
Packets injected behind the token and ahead of the token are considered separately.
If packets are injected only behind the token then the bound~\eqref{eqn:OF-RRW-no-jamming} on the length of a phase for \textsc{OF-RRW} applies, in that each phase takes at most $T=\frac{n}{1-\rho}$ rounds.
Such length of a phase is determined by the packets that are already queued when a phase begins.
Now, consider the effect of injections only ahead of the token while the old packets are already queued.
The duration of a phase is obtained from a duration $T$ of a phase of \textsc{OF-RRW} slowed down as much as possible by injecting packets in front of the token.
The upper bound on the duration of such a phase becomes
\begin{equation}
\label{eqn:RRW-no-jamming}
T(1+q + q^2+\cdots)=\frac{T}{1-\rho}\le \frac{n}{(1-\rho)^2}
\ .
\end{equation}
Packet latency is upper bounded by the duration of two consecutive phases.
The lengths of two consecutive phases are at most a sum of the lengths given by~\eqref{eqn:OF-RRW-no-jamming} and~\eqref{eqn:RRW-no-jamming}:
\[
\frac{n}{1-\rho} + \frac{n}{(1-\rho)^2} = \frac{2-\rho}{(1-\rho)^2}\cdot n
\ ,
\]
because injecting only in front of the token prevents creating old packets to be transmitted in the next phase, and the following phase starts with empty queues.
The second of these two phases may be additionally extended by at most  $\frac{\beta}{1-\rho}$,  due to the stretching-by-injecting effect, which gives the ultimate bound~\eqref{eqn:RRW-latency}.
\end{proof}

The bounds of Theorems~\ref{thm:RRW}  are asymptotically tight, which can be demonstrated by giving a specific adversary's strategy.
Let the adversary first keep injecting just after the token.
These packets are transmitted in the next phase, which simulates the behavior of \textsc{OF-RRW}.
Eventually the phase lengths gets arbitrarily close to $\frac{n}{1-\rho}$.
Then, at the beginning of a new phase, the adversary starts injecting just ahead of the token.
The duration of this one phase gets extended by an additional factor of $\frac{1}{1-\rho}$ due to stretching-by-injecting.

The tightness of the bounds implies that the advantage of the old-go-first mechanism applied in algorithm \textsc{OF-RRW}, as compared to \textsc{RRW}, is the speedup of packet latency by the following factor 
\[
\frac{2-\rho}{(1-\rho)^2} \cdot  \frac{1-\rho}{2} =\frac{1}{2}\cdot \frac{2-\rho}{1-\rho} > \frac{1}{2(1-\rho)}
\ ,
\] 
which is measured having an adversary fixed and $n$ growing unbounded.

\subsection{Channels with collision detection}

We consider algorithms \textsc{Old-First-Search-Round-Robin} (\textsc{OF-SRR}) and \textsc{Search-Round-Robin} (\textsc{SRR}), both of which use collision detection.
Executions are partitioned into phases.
A  phase denotes one full sweep of search through all the names of stations.

We begin with a technical estimate that will be used in proving bounds on packet latency.
Let $\lg x$ denote $\lceil \log_2 x \rceil$.


\begin{lemma}
\label{lem:phase-OF-SRR}

If there are already $x$ packets in the system when a phase of algorithm \textsc{OF-SRR} begins, then the phase takes at most $\min \,[ x\,(2+\lg n), x+2n-1]$ rounds.
\end{lemma}

\begin{proof}
We argue that there are at most $1+\lg n$ void rounds between two packets are heard on the channel.  
This is because of two reasons.
First, when a station finishes its transmissions, then one silent round either triggers the next search or completes the phase.
Second, when a new search to identify a station with a packet begins, it takes at most $\lg n$ collisions to identify a single station with pending packets.
There are also $x$ rounds spent to hear the $x$ packets. 

Next we give the following alternative estimate.
A phase can be represented by a binary search tree in which each interval on a stack corresponds to  a node.
In particular, a station with pending packets is in an interval that is a leaf, and an interval that creates a collision corresponds to an internal node. 
Observe that we may associate one void round with each node on such a tree.
The association depends on the kind of node.
First, if a node represents a station with packets, which is a leaf, then there is a silent round following all the transmissions by the station, which can be associated with the node.
Second, if this is an internal node, then it is associated with a collision.
It follows that the total number of nodes in the tree and the number of void rounds in a phase are equal. 
There are at most $2n-1$ nodes in the tree, because it has at most $n$ leaves.
The void rounds in the phase are added to the $x$ rounds used to hear the $x$ packets.
\end{proof}

Now we give the performance bounds for the algorithm \textsc{OF-SRR}.


\begin{theorem}
\label{thm:OF-SRR}

If algorithm \textsc{OF-SRR}  is executed by $n$ stations against an adversary of type $(\rho,\beta)$ then the number of packets simultaneously queued in the stations is at most
\begin{equation}
\label{eqn:OF-SRR-queues}
\frac{4\rho }{1-\rho}\cdot n  +\beta
\end{equation}
and packet latency is at most
\begin{equation}
\label{eqn:OF-SRR-latency}
\frac{4}{1-\rho}\cdot n +\beta(1+\rho)
\ .
\end{equation}
If $\rho\le\frac{1}{2+\lg n}$ then the number of packets simultaneously queued in the stations is at most $2\beta$ and packet latency is at most $2\beta(2+\lg n)$.
\end{theorem}

\begin{proof}
Let $T_i$ denote the duration of phase $i$, where $T_1=1$.
Let $Q_i$ denote the number of old packets in the beginning of phase~$i$, where $Q_1=0$.
Let $Q$ be an upper bound on the number of queued old packets and $T$ an upper bound on the duration of a phase.

First, we consider the case of $\rho\le \frac{1}{2+\lg n}$.
The inequality $Q_2\le \beta$ holds, including the effect of burstiness, so that $T_2\le \beta(2+\lg n)$.
Then again $Q_3\le \rho \cdot \beta (2+\lg n)\le \beta$.
The pattern repeats, so the invariants $Q_i\le \beta$ and $T_i\le  \beta(2+\lg n)$ are maintained.
This allows to set $Q= \beta$ and $T=  \beta(2+\lg n)$.
The queues size is at most the number of old and new packets together, which is $2Q=2\beta$, and packet latency is at most twice the duration of a phase~$T$, which is at most $2T=2\beta(2+\lg n)$.

Next, we consider the general case.
The sequences $(Q_i)_{i\ge 1}$ and $(T_i)_{i\ge 1}$ satisfy the following recursive dependencies, by Lemma~\ref{lem:phase-OF-SRR}, where we disregard the effect of burstiness:
\[
Q_{i+1}\le \rho\cdot T_i 
\]
and 
\[
T_{i+1}\le 2n+Q_{i+1}
\ .
\]
Iterating these recurrences produces the following bound $T$ on the duration of a phase:
\begin{equation}
\label{eqn:OF-SRR-no-jamming}
T_{i+1} \le 2n + \rho\cdot T_i \le 2n+ \rho 2n + \rho T_{i-1} \le 2n(1+\rho+\rho^2+\ldots)\le \frac{2n}{1-\rho}=T
\ .
\end{equation}
A packet spends at most two consecutive phases waiting to be heard, each phase taking at most~$T$ rounds.
A bound for $T$ given in~\eqref{eqn:OF-SRR-no-jamming} disregards the effect of burstiness, which can be accounted for as follows.
If the adversary injects $\beta$ packets in one round then this increases the number of packet queued in the current phase. 
This injection extends the duration of the next phase rather then the current one, because this will  be the phase when these packets are transmitted as old.
These extra transmissions make it possible for the adversary to inject $\rho \beta$ packets, which extends the duration of the next phase by $\rho \beta$ rounds.

Here are the concluding estimates.
The maximum number of queued packets is at most $\rho T$ old packets added to at most $\rho T$ new packets, and at most $\beta$ packets injected in a burst, which gives~\eqref{eqn:OF-SRR-queues}.
The maximum number of rounds spent by a packet waiting to be heard on the channel is twice the upper bound~$T$ on a duration of a phase~\eqref{eqn:OF-SRR-no-jamming}, incremented by $\beta$ extra rounds in a phase of a bursty injection along with $\rho\beta$ rounds of the next phase, which gives~\eqref{eqn:OF-SRR-latency}.
\end{proof}

The bounds of Theorems~\ref{thm:OF-SRR} are asymptotically tight, which can be shown as follows.
There are two bounds on queues and latency, and tightness of a bound occurs when the adversary's type satisfies additional conditions. 
First, the case of small $\rho$, say, $\rho=\frac{1}{2\lg n}$.
Queues size is tight as the bound is proportional to the burstiness component.
Let the adversary inject packets in pairs into two adjacent stations, a packet per station, such that they are at some point together in an interval on the stack that is of a constant-size.
There are $\beta/2$ such pairs, and they are injected into stations that are about $2n/\beta$ apart.
For the burstiness component~$\beta$ such that $\lg \beta =o(\lg n)$, it takes $\Omega(\beta\log n)$ to transmit $\beta$ packets injected simultaneously, so the phase duration is also tight.
Next, the case of large injection rates~$\rho$, in particular, when $\rho>\frac{1}{2}$.
Let the adversary keep injecting into pairs of stations, a packet per station, that belong together to intervals of a constant length that are on the stack at some point in time. 
The time spent waiting to hear a new packets is $\Omega(\log n)$ initially, while the adversary injects at a rate larger than $\frac{1}{2}$.
Eventually, the rate of hearing consecutive packets becomes $\Omega(1)$, but at that point the number of packets queued becomes $\Omega(\rho n)$. 
The adversary continues injecting at full power to extend a phase's length close to $\Omega(\frac{n}{1-\rho})$, by the stretching-by-injecting effect.
The adversary may add $\beta$ packets in a burst and next extend the two following phases by about $\beta+\rho\beta$ rounds. 

Next, we consider algorithm \textsc{SRR}.
We begin with a preliminary fact.


\begin{lemma}
\label{lem:phase-SRR}

Let us consider the beginning of a phase of algorithm \textsc{SRR}.
If the number of packets that are either already queued  or they are injected during the phase into stations that belong to some intervals on the stack is~$y$ then the phase takes at most $\min ( y\,(2+\lg n), y+2n-1)$ rounds.
\end{lemma}

\begin{proof}
A proof is similar to that of Lemma~\ref{lem:phase-OF-SRR}, with a difference regarding  which packets get transmitted in a current phase.
While in algorithm \textsc{OF-SRR} these are the packets already queued when the phase starts, algorithm \textsc{SRR} has all the available packets transmitted, including those already present when the phase begins but also newly injected ones.
Each station that holds packets competes for access to the channel in a phase, unless its name is no longer on an interval on the stack.
A round of the first transmission by such a station occurs when the interval including the station's name is removed from the top of the stack and the station is the only one in the interval that holds packets pending transmission.
\end{proof}

Now we give the performance bounds for the algorithm \textsc{SRR}.


\begin{theorem}
\label{thm:SRR}

If algorithm \textsc{SRR}  is executed by $n$ stations against an adversary of type $(\rho,\beta)$ then the number of packets simultaneously queued  is at most
\begin{equation}
\label{eqn:SRR-queues}
\frac{4\rho }{1-\rho}\cdot n +\beta
\end{equation}
and packet latency is at most
\begin{equation}
\label{eqn:SRR-latency}
\frac{4-2\rho}{(1-\rho)^2}\cdot n +\frac{\beta}{1-\rho} 
\ .
\end{equation}
If $\rho\le\frac{1}{2+\lg n}$ then the number of packets simultaneously queued is at most $2\beta$ and packet latency is at most $3\beta(2+\lg n)$.
\end{theorem}

\begin{proof}
Packets that are injected into stations that do not belong to the intervals on the stack are transmitted in the next phase.
The way the algorithm handles these packets is consistent with the design of \textsc{OF-SRR}, so their packet latency conforms to the bound on packet latency for \textsc{OF-SRR}.
Packets injected into stations that belong to the intervals on the stack are transmitted in the current phase, which may slow down the phase as compared to \textsc{OF-SRR}.
The extra rounds are either spent on transmissions or they produce collisions while a next station with packets is identified.
Each round spent on transmissions decreases the number of packets in the queues, but not each of these rounds is used by the adversary to inject new packets and so increase the number of packets queued.
Regarding the rounds producing collisions, they are estimated as overheads of either  $2+\lg n$ per packet or $2n-1$ total in a phase, but in this respect Lemma~\ref{lem:phase-SRR} has exactly the same overheads as Lemma~\ref{lem:phase-OF-SRR}.
The upper bound on the number of packets stored in the queues derived for \textsc{OF-SRR} includes both the old and new packets, but accounted for separately.
Since accounting for transmission of old and new packets together is consistent with accounting for them separately, the upper bounds on the size of queues of \textsc{OF-SRR} also applies to algorithm~\textsc{SRR}.
We conclude that the bound \eqref{eqn:SRR-queues} on queues size can be made identical to \eqref{eqn:SRR-queues}, along  with the bound of~$2\beta$ for the suitably small injection rates.

Next we estimate packet latency.
There are two cases, the general one and a special one of suitably small injection rates.
Let $T_i$ denote the duration of phase $i$, and $T$ an upper bound on the duration of a phase.

First the case of $\rho\le \frac{1}{2+\lg n}$.
If the adversary injects only into stations that are not on the stack then these packets are old, in the sense that they will be heard in the next phase, so the bound $T=  \beta(2+\lg n)$ for algorithm \textsc{OF-SRR} applies.
If the adversary injects only into stations that are still on the stack, then this allows to extend a phase's duration by a factor of $1/(1-\rho)$. 
A packet can be delayed at most two consecutive phases, which is the following, for sufficiently large~$n$:
\[
T\Bigl(1+\frac{1}{1-\rho}\Bigr) 
= T\cdot \frac{2-\rho}{1-\rho}
\le T\cdot\frac{2}{1-\frac{1}{2+\lg n}}
= 2T\cdot\frac{2+\lg n}{1+\lg n}
\le 3T
\ .
\]

Next, we consider the general case.
If packets are injected only into stations that do not belong to the intervals on the stack at the round of injection, then the bound~\eqref{eqn:OF-SRR-no-jamming} on the length of phase for \textsc{OF-SRR} applies, in that a phase takes at most $T=\frac{2n}{1-\rho}$ rounds.
Each such a duration suffices to hear the packets that are already queued when a phase begins.
Now, consider the effect of injections only into stations that belong to intervals on the stack at the round of injection, while the old packets are already queued.
The duration of a phase is obtained from the duration~$T$ of a phase of \textsc{OF-SRR} slowed down as much as possible by injecting packets into stations whose names are in the intervals on the stack.
An upper bound on the duration of such a phase is obtained by the stretching-by-injecting effect to be at most the following:
\begin{equation}
\label{eqn:phase-SRR-no-jamming}
T(1+\rho + \rho^2+\cdots)=\frac{T}{1-\rho}\le \frac{2n}{(1-\rho)^2}
\ .
\end{equation}
The maximum of a sum of lengths of two consecutive phases is obtained as a sum of the lengths given by~\eqref{eqn:OF-SRR-no-jamming} and~\eqref{eqn:phase-SRR-no-jamming}, because injecting only into stations on the stack results in not creating any old packets to be heard in the next phase.
The obtained bound is as follows:
\[
\frac{2n}{1-\rho} + \frac{2n}{(1-\rho)^2} = \frac{4-2\rho}{(1-\rho)^2}\cdot n
\ .
\]
The second of these two phases may be additionally extended by at most  $\frac{\beta}{1-\rho}$,  due to the stretching-by-injecting effect combined with burstiness, which gives~\eqref{eqn:SRR-latency}.
\end{proof}

The bounds of Theorem~\ref{thm:SRR} are asymptotically tight, which can be shown by finding a specific adversary's strategy.
The case of small injection rate is similar as for algorithm \textsc{OF-SRR}, since algorithm \textsc{SRR} has its performance bounds differ from those for \textsc{OF-SRR} by constant multiplicative factors when injection rates are smaller than $1/(2+\lg n)$.
Next we discuss the general case.
Let the adversary first keep injecting into stations whose names are not in the intervals on the stack, similarly as in the case of algorithm \textsc{OF-SRR}.
These packets are transmitted in the next phase, which is consistent with the behavior of \textsc{OF-SRR}, so that eventually the phase lengths gets arbitrarily close to $\frac{2n}{1-\rho}$.
Then, at the beginning of a new phase, the adversary starts injecting into stations that are still on the stack.
The duration of this one phase can get extended by an additional factor of $\frac{1}{1-\rho}$ due to  stretching-by-injecting.
This same phase can be further extended by $\frac{\beta}{1-\rho}$ by burstiness amplified by stretching-by-injecting.

The tightness of the bounds implies that the advantage of the old-go-first mechanism applied in algorithm \textsc{OF-SRR}, as compared to \textsc{SRR}, is the speedup of packet latency by a factor that is  grater than $\frac{1}{2(1-\rho)}$, similarly as in the case of algorithm \textsc{OF-RRW} compared to \textsc{RRW}.

\section{An Adaptive Algorithm without Jamming}

\label{sec:adaptive-algorithms-without-jamming}

Algorithm \textsc{Move-Big-To-Front} (\textsc{MBTF}) is an adaptive one for channels without collision detection.
This algorithm is stable even when injection rate is~$1$, but for this rate packet latency is unbounded, in that even an eventual hearing of a packet is not guaranteed~\cite{ChlebusKR09}.

Algorithm \textsc{MBTF} works with stations arranged in a dynamic list, and we refer to the stations not by their names but by their positions on this list.
There are $n$ positions: $1,2,3,\ldots, n$, with station~$1$ at the front of the list  and station~$n$ at the end.

The list of stations is traversed by a token that gives the right to transmit.
Let a traversal of the token, which starts at the front of the list and ends by reaching again the front station of the list, be called a \emph{pass} of the token.
A pass is concluded by either discovering a new big station or traversing the list to its end.  

We monitor the number of packets in the queues at the end of a pass, to see how the pass contributed to  the number of packets stored in the queues.
If the number of queued packets  at the end of a pass is smaller than at the end of the previous pass, then such a pass  is called \emph{decreasing}, otherwise it is \emph{non-decreasing}.

We partition passes into two categories, depending on whether a big station is discovered in a pass or not.
If a big station is discovered in a pass then such a pass is called \emph{big} and otherwise it is called \emph{small}.
A discovery of a big station results in moving this big station to the front of the list, which concludes  the pass.
The next pass begins by a transmission of the newly discovered big station, just after it is  moved up to the front position in the list.
We begin the analysis of performance of algorithm \textsc{MBTF} by investigating how many packets can be accumulated in the queues when small passes occur.


\begin{lemma}
\label{lem:queues-MBTF-small}

If algorithm \textsc{MBTF} is executed by $n$ stations against an adversary of type $(\rho,\beta)$, in such a manner that all the passes have been small up to a given round, then the number of packets stored in the queues in this round is at most $\rho  n^2 + \beta$.
\end{lemma}

\begin{proof}
If the adversary injects packets at the rate as close to injection rate as possible then burstiness component can be applied only once, and we will conclude with its contribution, while initially we disregard it.
A small pass takes $n$ rounds.
The adversary can inject $\rho n$ packets during a time segment of these many rounds. 
This number $\rho n$ is also an upper bound on the number of stations with packets during a non-decreasing small pass, because if there were more such stations, then each of them would transmit a packet during a pass.

Each station with packets has at most $n-1$ packets during a small pass.  
It follows that if a small pass is non-decreasing then the number of packets in the queues at the end of the pass is at most $\rho n\cdot(n-1)$.
The adversary can inject at most $\rho n +\beta$ packets in the course of any of these passes.
We conclude that the number of packets is at most $\rho  n^2 + \beta$ in a round by which only  small passes have occurred.
\end{proof} 

The adversary may use big passes to accumulate packets in queues and delay  packets at the end of the list of stations by preventing the token to reach the tail of this list.
The accumulation of packets is largest when the token traverses as many stations with empty queues as possible before discovering a big station.
During such passes, the adversary can inject at the rate of~$\rho$ while striving to make the ratio of the number of rounds with messages heard on the channel smaller than~$\rho$, which results in the number of queued packets growing.


\begin{theorem}
\label{thm:MBTF}

If algorithm  \textsc{MBTF} is executed by $n$ stations against an adversary of type $(\rho,\beta)$  then the number of packets stored in the queues  in any round is at most 
\begin{equation}        
\label{eqn:MBTF-packets}
\rho \,(1+\rho)\,n^2 + \beta 
\end{equation}
and packet latency is at most
\begin{equation}
\label{eqn:MBTF-latency}
\frac{1+\rho -\rho^2}{1-\rho} \cdot n^2 + \frac{\beta}{1-\rho} 
\ .
\end{equation}
\end{theorem}

\begin{proof}
We will disregard the burstiness component through the initial stages of the analysis, to apply it at the end of the process of accounting for time and injected packets.

By Lemma~\ref{lem:queues-MBTF-small},  if no big station has been discovered yet then there are at most  $\rho n^2$ packets in total.
We explore now how much the queues can increase when big passes occur.
If there are at most~$\rho n$ stations with packets then the sum of the lengths of big passes is maximized when the following is the case: (1)~stations holding packets are located at the end of the list, and   (2)~each time the token reaches one of these stations, for the first time since big passes started to occur, then the station is discovered to be big.
Therefore, the sum of the lengths of big passes  is at most the following:
\[
\sum_{i=1}^{\rho n} \bigl(n-\rho n+i\bigr)
\le
\rho n^2 
\ ,
\]
for sufficiently large $n$.
During these big passes, at most $\rho\cdot \rho n^2$ new packets are injected. 
The total number of packets at this point is at most
\[
\rho n^2 + \rho^2 n^2 = \rho (1+\rho) \cdot n^2
\ .
\]
Injecting $\beta$ packets in one round can be increase the total number of packets to at most~\eqref{eqn:MBTF-packets}.

Next we estimate packet latency.
Let us consider some packet $p$ and we argue about its delay by building a worst-case scenario.
We may assume that $p$ gets injected when the configuration of packets is already as in Lemma~\ref{lem:queues-MBTF-small}, which is such that at most $\rho  n^2 $ packets are located  in the $\rho  n$ stations located at the end of the list, each holding at most  $n$ packets, but possibly fewer.
Let  packet~$p$ be injected into the last station, which takes the longest for the token to reach when starting from the front.
Additionally, if the last station is never discovered to be big, which is the case when the total number of packets in this station is at most $n-1$ including~$p$, then the token will never discover the station to be big before a packet that is at the bottom when $p$ is injected is ready to be transmitted.
Packet~$p$ may be at the bottom of its queue just after it is injected, and we may assume it is preceded  by $n-2$ packets in its queue.
The token will need to cover the whole length of the list $n-1$ times to reach~$p$ when it is already ready to be transmitted.
Each such a traversal of the whole list makes a small pass. 
In the meantime, the token may be delayed by discovering big stations, what makes the token return back to the front station without reaching the station holding~$p$.

We estimate how much time may pass before the token finally visits the $p$'s station, when $p$ is already  at the top of the queue ready to be transmitted, by accounting for the following three groups of rounds contributing to $p$'s waiting time: 
\begin{enumerate}
\item[(1)] a delay due to discovering big stations, 

\item[(2)] a delay due to small passes and packets injected during such passes, 

\item[(3)] the effect of burstiness.
\end{enumerate}
We begin with the effect of discovering big stations.
Starting from the $p$'s injection, the adversary may inject packets into the trailing $\rho n$ stations to make each of them big, with the exception of the last one.
The discoveries of up to $\rho  n$ big stations at the end of the list provide delays of up to these many rounds:
\[
\sum_{i=1}^{\rho n-1} \bigl(n-\rho n+i\bigr)
\le
\rho n^2 
\ .
\]
During these big passes, a worst-case waiting scenario occurs when they are extended by stretching-by-injecting to at most these many rounds in total:
\[
\frac{\rho }{1-\rho} \cdot n^2
\ .
\]

Next, we consider the effect of small passes.
It has two components.
There are $n-1$  small passes before the token reaches~$p$ when at the top of its queue, each pass contributing $n$ rounds, for the total of $n(n-1)\le n^2$ rounds, which is the first component.

During small passes,  packets can be injected to introduce additional delay, possibly through discovering big stations.
Suppose some $x$ such packets are injected.
If they are located in big stations that are discovered big for the first time then there are at most~$\frac{x}{n}$ such stations, each contributing a delay of at most~$n$ rounds for the total of at most $\frac{x}{n}\cdot n=x$ rounds of delay.
Otherwise, if some new packets are injected into a station that has already been discovered big and is at position~$i$ in the list, then this station has at most $n-i$ packets inherited from the time it was discovered big and moved to the front, so at least $i$ packets are needed to make it big again,  and these $i$ packets contribute to delay~$i$ by making the station big.
Any excess of $y$ packets beyond $n$ injected into a big station will contribute to a delay of~$y$ when the station is moved to the front of the list and starts transmitting.
So overall, the delay is upper bounded by the number of packets injected.
There are at most $\rho\cdot n^2$ packets injected during small passes.
The resulting delay is at most such, which is the second component.

Finally, burstiness allows to inject $\beta$ packets into a big station, which can be extended to $\frac{\beta}{1-\rho}$ by stretching-by-injecting.

We have assessed the three contributions to packet delay.
Adding them together gives a total of at most these many rounds:
\[
 \frac{\rho }{1-\rho} \cdot n^2 +  (1+\rho) \cdot n^2 +\frac{\beta}{1-\rho}
 =  
 \frac{\rho+(1+\rho)(1-\rho)}{1-\rho} \cdot  n^2 +\frac{\beta}{1-\rho}
=
\frac{1+\rho-\rho^2}{1-\rho} \cdot n^2+\frac{\beta}{1-\rho}
\ ,
\]
which is the claimed upper bound on packet latency~\eqref{eqn:MBTF-latency}. 
\end{proof}

The bounds given in Theorem~\ref{thm:MBTF} are asymptotically tight.
The factors $1+\rho$ and $1+\rho-\rho^2$ in the upper bounds~\eqref{eqn:MBTF-packets} and~\eqref{eqn:MBTF-latency} are~$\Theta(1)$ because $1<1+\rho-\rho^2< 2$.
It is sufficient to show how to construct a configuration with $\Omega(\rho n^2 +\beta)$ queued packets and a packet whose delay is  $\Omega\bigl(\frac{n^2 + \beta}{1-\rho}\bigr)$.

Let the adversary build queues of $n-1$ packets each in $\frac{\rho n }{2}$ stations.
This occurs in the course of small passes during which the adversary injects two packets into each of some fixed $\frac{\rho n }{2}$ stations, so each of them grows in a pass.
After $n-1$ such small passes, each of the stations with packets has $n-1$ packets.
During one more pass, the adversary injects $\rho n + \beta$ packets so that the number of queued packets is at least $\frac{\rho n^2 }{2} +\beta$.

Next we consider packet latency.
Let the adversary build queues of $n-1$ packets each in the last  $\frac{\rho n}{2}\le \frac{n}{2}$ stations, while the first $\frac{1-\rho }{2} \cdot n\ge \frac{n}{2}$ stations have empty queues.
A packet $p$ is injected into the last station as its last packet at the bottom.
Let the adversary make each of the stations with packets big by inserting one extra packet, starting with the station in the smallest position, but skipping the last station.
After that, the adversary keeps injecting at full power into the station that is last but one, which also includes  injecting $\beta$ packets in one round.

We consider two cases.
The first case is when $\rho\le \frac{1}{2}$, which implies $\frac{n^2 +\beta}{1-\rho}\le 2 n^2 + 2\beta$.
The big passes contribute at least~$\beta$ rounds and the small passes that follow contribute at least~$n(n-2)$ rounds.
The second case is when $\rho > \frac{1}{2}$.
The number of void rounds in big passes is at least  
$\sum_{i=1}^{\frac{n}{4}} \bigl(n-\frac{n}{2} +i\bigr) \ge \frac{n^2}{8}$.
When the last-by-one station is discovered big, the adversary injects additional~$\beta$  packets into it.
The number of rounds $\frac{n^2}{8}+\beta$ can be extended by stretching-by-injecting to at least $\frac{1}{8} \cdot \frac{n^2+\beta}{1-\rho}$ rounds.
All these rounds contribute to the delay of packet~$p$.
This quantity grows unbounded if injection rate~$\rho$ converges to~$1$.

\section{Non-adaptive Algorithms with Jamming}

\label{sec:non-adaptive-algorithms-with-jamming}

We show that non-adaptive algorithms may have bounded worst-case packet latency on channels with jamming.  
The caveat is that they are correct only  against adversaries whose jamming burstiness is bounded from above by a parameter we denote~$J$.
This parameter $J$ is part of code, and to emphasize this, is included as part of the names of algorithms \textsc{OF-JRRW($J$)} and \textsc{JRRW($J$)}.
The value of~$J$ does not occur in the upper bounds on packet latency we derive, as the jamming burstiness of a jamming adversary of type $(\rho,\lambda,\beta)$ is at most $\beta/(1-\lambda)$.


\begin{lemma}
\label{lem:cycle-plus-packets}

If there are $x$ old packets in the queues when a phase of algorithm \textsc{OF-JRRW($J$)} executed by $n$ stations begins, against an adversary of type $(\rho,\lambda,\beta)$ whose jamming burstiness is at most~$J$, then the phase takes at most these many rounds:
\[
\frac{x + n(J+1) + \beta}{1-\lambda}
\ .
\] 
\end{lemma}

\begin{proof}
It takes $x$ rounds to transmit the $x$ old packets.
It takes $n$ intervals, of $J+1$ void rounds each, for the token to make a full cycle and so  visit every station with old packets.
Therefore, at most $n(J+1)+x$ clear rounds are needed to hear the $x$ old packets. 
Consider a contiguous time segment of $z$ rounds in which some $x$ packets are heard. 
At most $z \lambda + \beta$ of these $z$ rounds can be jammed. 
Therefore, the following inequality needs to hold:
\[
z \le   n (J+1) + x + z \lambda + \beta
\ .
\] 
Solving for $z$, we obtain the following bound
\[
z \le \frac{x + n(J+1)+\beta}{1-\lambda}
\]
on a length of a contiguous time interval in which at least $x$ packets are heard.
\end{proof}

Lemma~\ref{lem:cycle-plus-packets} could be explained by referring to the stretching-by-jamming effect directly: there are $x$ rounds to successfully transmit the old packets, there are $n (J+1)$ rounds to get the token around, and there is the burstiness component $\beta$, each of them stretched by the factor $1/(1-\lambda)$.
A phase takes close to the upper bound in Lemma~\ref{lem:cycle-plus-packets} when the adversary does not jam the $n$ intervals of $J+1$ void rounds, each used to advance the token once.
In what follows, similar facts are argued about by referring directly to the stretching-by-jamming effect.

During analyses of algorithms, if rounds are counted in disjoint intervals and the adversary jams at full power then the burstiness component can be applied only once.
So Lemma~\ref{lem:cycle-plus-packets} may be used for one phase  as formulated above, and in the remaining ones the bound is restricted to a smaller quantity $(x + n(J+1))/(1-\lambda)$.


\begin{theorem}
\label{thm:OF-JRRW(J)-jamming}

If algorithm \textsc{OF-JRRW($J$)} is executed by $n$ stations against a jamming adversary of type $(\rho,\lambda,\beta)$ such that its jamming burstiness at most~$J$ then the number of packets  queued in any round is at most
\begin{equation}
\label{eqn:OF-JRRW(J)-queues}
\frac{2(\beta+1)}{1-\rho-\lambda} \cdot n+\beta
\end{equation}
and packet latency  is  at most 
\begin{equation}
\label{eqn:OF-JRRW(J)-latency}
\frac{2(\beta+1)}{(1-\lambda)(1-\rho-\lambda)} \cdot n + \frac{\beta(1+\rho-\lambda)}{(1-\lambda)^2}    
\ .
\end{equation}
\end{theorem}

\begin{proof}
Let $T_i$ be the duration of phase~$i$ and $Q_i$ be the number of old packets in the beginning of phase~$i$, for $i\ge 1$.
The following two estimates lead to a recurrence for the numbers~$T_i$, in which we disregard the burstiness component.
One estimate reads  
\begin{equation}
\label{eqn:C-MBTF-lemma-non-adaptive-algorithms-with-jamming}
Q_{i+1}\le \rho T_i \ , 
\end{equation}
by the definitions of old packets and of type $(\rho,\lambda,\beta)$ of the adversary, and the other estimate is 
\begin{equation}
\label{eqn:two-non-adaptive-algorithms-with-jamming}
T_{i+1}\le \frac{n(J+1)+Q_{i+1}}{1-\lambda} \ ,
\end{equation}
by Lemma~\ref{lem:cycle-plus-packets}.  
Let us denote $n(J+1)=a$.
Substitute \eqref{eqn:C-MBTF-lemma-non-adaptive-algorithms-with-jamming} into \eqref{eqn:two-non-adaptive-algorithms-with-jamming} to obtain
\[
T_{i+1} 
\le
\frac{a + Q_{i+1}}{1-\lambda} 
\le 
\frac{a}{1-\lambda}+ \frac{\rho }{1-\lambda}  \cdot T_i
\le c + d T_i
\ ,
\]
for $c=\frac{a}{1-\lambda}$ and $d=\frac{\rho}{1-\lambda}$.
Note that $d < 1$, as $\rho<1-\lambda$.
An upper bound on the duration of a phase is found by iterating the recurrence $T_{i+1}\le c + d T_i$ to obtain a bound on the duration $T$ of a phase:
\begin{equation}
\label{eqn:three-non-adaptive-algorithms-with-jamming}
c+d c + d^2 c +\ldots d^i c\le \frac{c}{1-d}=T\ .
\end{equation}
After substituting $c=\frac{a}{1-\lambda}$ and $d=\frac{\rho}{1-\lambda}$ into~\eqref{eqn:three-non-adaptive-algorithms-with-jamming}, we obtain the following estimate:
\begin{equation}
\label{eqn:four-non-adaptive-algorithms-with-jamming}
T =
\frac{a}{1-\lambda} \cdot \frac{1}{1-\frac{\rho}{1-\lambda}}
=
\frac{a}{1-\rho-\lambda} \ .
\end{equation}
Replacing $a$ by $n(J+1)$ in~\eqref{eqn:four-non-adaptive-algorithms-with-jamming}  expands $T$ to the following quantity:
\begin{equation}
\label{eqn:five-non-adaptive-algorithms-with-jamming}
T =
\frac{n(J+1)}{1-\rho-\lambda} \ .
\end{equation}
We apply the estimate $J\le \beta/(1-\lambda)$ to~\eqref{eqn:five-non-adaptive-algorithms-with-jamming} to obtain the following upper bound on $T$:
\begin{equation}
\label{eqn:nine-non-adaptive-algorithms-with-jamming}
T \le 
\frac{n(\frac{\beta}{1-\lambda}+1)}{1-\rho-\lambda}
=
\frac{n(\beta+1-\lambda)}{(1-\lambda)(1-\rho-\lambda)}
\le
\frac{n(\beta+1)}{(1-\lambda)(1-\rho-\lambda)} \ .
\end{equation}

A packet waits to be transmitted through at most two consecutive phases, each taking at most~$T$ rounds, where a bound for~$T$ given in~\eqref{eqn:nine-non-adaptive-algorithms-with-jamming} does not account for burstiness.
Let the adversary inject extra $\beta$ packets in a round of a phase.
This increases the number of packets in the current phase but extends the duration of the next phase by $\frac{\beta}{1-\lambda}$, which is the phase when these packets are transmitted as old.
These transmissions in turn allow the adversary to inject $\rho\cdot \frac{\beta}{1-\lambda}$ additional packets, which extends the duration of the immediately following phase by $ \frac{\rho\beta}{(1-\lambda)^2}$ rounds by the stretching-by-jamming effect.

We conclude with the following estimates.
The maximum number of queued packets is obtained by adding at most $\rho T$ old packets to at most $\rho T$ new packets, along with at most $\beta$ packets injected in a burst, which together makes the following bound:
\[
\frac{2\rho n(\beta+1)}{(1-\lambda)(1-\rho-\lambda)}+\beta \le\frac{2n(\beta+1)}{1-\rho-\lambda}+\beta
\ ,
\]
where we used $\rho< 1-\lambda$.
This yields~\eqref{eqn:OF-JRRW(J)-queues}.
The maximum number of rounds spent by a packet waiting to be heard on the channel is obtained by adding twice the upper bound~$T$ on a duration of a phase~\eqref{eqn:nine-non-adaptive-algorithms-with-jamming}, incremented by $\frac{\beta}{1-\lambda}$ extra rounds in the phase immediately following one of a bursty injection, along with $ \frac{\rho\beta}{(1-\lambda)^2}$ rounds  of the following phase.
This gives the following amount:
\[
2\cdot \frac{n(\beta+1)}{(1-\lambda)(1-\rho-\lambda)} + \frac{\beta}{1-\lambda} + \frac{\rho\beta}{(1-\lambda)^2}
=
\frac{2n(\beta+1)}{(1-\lambda)(1-\rho-\lambda)} +\frac{\beta(1+\rho-\lambda)}{(1-\lambda)^2}
\ ,
\]
where we used $\rho< 1-\lambda$.
This yields~\eqref{eqn:OF-JRRW(J)-latency}.
\end{proof}

The bound of Theorem~\ref{thm:OF-JRRW(J)-jamming}  is tight, by the following scenario.
A phase includes $n(J+1)$ void rounds to advance the token around, which the adversary does not jam.
If the adversary injects at full power, and at the same time jams  at full power the rounds during which some station tries to transmit, then this is equivalent to injections with rate $\frac{\rho}{1-\lambda}$.
Eventually phases get arbitrarily close to the following magnitude, by combined stretching:
\[
\frac{n(J+1)}{1-\lambda} \cdot \frac{1}{1-\frac{\rho}{1-\lambda}} =\frac{n(J+1)}{1-\rho-\lambda}
\ .
\]
If the adversary is such that $J=\frac{\beta}{1-\lambda}$ then a phase takes close to $\frac{n\beta}{(1-\lambda)(1-\rho-\lambda)}$ rounds.
The number of packets injected during a phase of such duration can be made close to
$\rho\cdot \frac{ n\beta}{(1-\lambda)(1-\rho-\lambda)}$, which can be made asymptotic  to $\frac{n\beta}{1-\rho-\lambda}$, if $\rho=\Theta(1-\lambda)$.

Next, we analyze algorithm \textsc{JRRW}($J$).


\begin{theorem}
\label{thm:JRRW(J)-jamming}

If algorithm \textsc{JRRW($J$)} is executed by $n$ stations against a jamming adversary of type $(\rho,\lambda,\beta)$ such that its jamming burstiness at most~$J$ then the number of packets stored in the queues in any round is at most
\begin{equation}
\label{eqn:JRRW(J)-jamming-queues}
\frac{2(\beta+1)}{1-\rho-\lambda} \cdot n +\beta
\end{equation}
and packet latency  is  at most 
\begin{equation}
\label{eqn:JRRW(J)-jamming-latency}
\frac{2(\beta+1)}{(1-\lambda)(1-\rho-\lambda)^2} \cdot n + \frac{\beta(1-\lambda)}{1-\rho-\lambda}
\ .
\end{equation}
\end{theorem}

\begin{proof}
Packets injected by the adversary may be transmitted in the current phase or in the next one, depending one how the station into which they are injected is related to the station with a token.
We consider separately the impact of such injections to extend phases, by first estimating the phase length when packets are transmitted in the next phase and then when they are transmitted in the current phase.

Packets injected at stations behind the one that holds the token at the moment are transmitted in the next phase.
These new packets will be visited by the token only after they become old.
It follows that the adversary can make algorithm \textsc{JRRW}($J$) behave as \textsc{OF-JRRW$(J)$} by  choosing stations to inject packets into in this very manner.
If all packets are injected this way, an upper bound on the duration of a phase is given by~\eqref{eqn:nine-non-adaptive-algorithms-with-jamming}, which we denote by~$T=\frac{n(\beta+1)}{(1-\lambda)(1-\rho-\lambda)}$.

Next, we estimate the contribution of packets injected at stations ahead of the station that holds the token at the moment, and which are transmitted in the current phase, compounded with packets already at the stations,  which were injected behind the station holding the token.
The packets get injected with the rate extended by stretching-by-jamming effect.
The total number of rounds in such a phase is at most
\begin{equation}
\label{eqn:seven-non-adaptive-algorithms-with-jamming}
T + T\cdot \frac{\rho}{1-\lambda}+ T \cdot \Bigl(\frac{\rho}{1-\lambda}\Bigr)^2  +\ldots 
=  
\frac{T}{1-\frac{\rho}{1-\lambda}}
=
T\cdot\frac{1-\lambda}{1-\rho-\lambda}
\ .
\end{equation}
Substituting $T=\frac{n(\beta+1)}{(1-\lambda)(1-\rho-\lambda)}$ into~\eqref{eqn:seven-non-adaptive-algorithms-with-jamming} results in the following bound
\begin{equation}
\label{eqn:ten-non-adaptive-algorithms-with-jamming}
\frac{n(\beta+1)}{(1-\lambda)(1-\rho-\lambda)}\cdot \frac{1-\lambda}{1-\rho-\lambda}
=
\frac{n(\beta+1)}{(1-\rho-\lambda)^2}
\ ,
\end{equation}
which is the maximum possible length of a single phase, if we disregard the effects of burstiness.
To account for burstiness, the adversary can inject $\beta$ packets in front of the token, and then by iterating stretching-by-jamming by injecting at full power, the resulting extra $\beta$ rounds get extended to $\beta \cdot \frac{1-\lambda}{1-\rho-\lambda}$.
The duration of two consecutive phases is bounded from above by a sum of \eqref{eqn:nine-non-adaptive-algorithms-with-jamming}, which we denote by~$T=\frac{n(\beta+1)}{(1-\lambda)(1-\rho-\lambda)}$, of~\eqref{eqn:ten-non-adaptive-algorithms-with-jamming}, and of a one-time extension of a phase due to burstiness, which we calculated to be~$\beta \cdot \frac{1-\lambda}{1-\rho-\lambda}$.
They together make the following bound:
\[
\frac{n(\beta+1)}{(1-\lambda)(1-\rho-\lambda)} + \frac{n(\beta+1)}{(1-\rho-\lambda)^2} + \beta \cdot \frac{1-\lambda}{1-\rho-\lambda}
\le
\frac{2n(\beta+1)}{(1-\lambda)(1-\rho-\lambda)^2} + \frac{\beta(1-\lambda)}{1-\rho-\lambda}
\ ,
\]
which is the upper bound~\eqref{eqn:JRRW(J)-jamming-latency}.
\end{proof}

The upper bound given in Theorem~\ref{thm:JRRW(J)-jamming} is asymptotically tight, which can be justified by the following scenario.
Let the adversary initially inject behind the token, which results in all injected packets transmitted in the next phase.
The accompanying pattern of jamming is such as to make queues and packet latency get asymptotic to the bounds given in Theorem~\ref{thm:OF-JRRW(J)-jamming}.
This gives the tightness of queue bounds, as they are identical in Theorems~\ref{thm:OF-JRRW(J)-jamming} and~\ref{thm:JRRW(J)-jamming}.
At this point, a phase takes close to $\frac{n\beta}{(1-\lambda)(1-\rho-\lambda)}$ rounds.
Now, the adversary switches to injecting just before the token, to make the old packets injected in the previous phase and the currently injected packets transmitted in the current phase, so there are no outstanding packets when the phase is over.
Injecting and jamming at full power has the effect of stretching injection rate to $\frac{\rho}{1-\lambda}$, which eventually makes a phase take close to the following amount: 
\[
\frac{n\beta}{(1-\lambda)(1-\rho-\lambda)} \cdot \frac{1-\lambda}{1-\rho-\lambda}
=
\frac{n\beta}{(1-\rho-\lambda)^2}
\ ,
\]
by the estimate as in~\eqref{eqn:seven-non-adaptive-algorithms-with-jamming}, which is asymptotic to~\eqref{eqn:JRRW(J)-jamming-latency}, if $\lambda=\Theta(1)$.

The upper bound on packet latency given in Theorem~\ref{thm:JRRW(J)-jamming} differs by the factor $\frac{1}{1-\rho-\lambda}>1$ from the bound in Theorem~\ref{thm:OF-JRRW(J)-jamming}.
This factor can become arbitrarily large when $\rho+\lambda$ gets suitably close to~$1$. 
This difference between the two bounds reflects the benefit of the approach ``old-go-first'' applied in the design of algorithm \textsc{OF-JRRW$(J)$}, as compared to algorithm \textsc{JRRW$(J)$}.

\section{Adaptive Algorithms with Jamming}

\label{sec:adaptive-algorithms-with-jamming}

We give worst-case upper bounds on queues size and packet latency against jamming adversaries for the following three adaptive algorithms: \textsc{OFC-RRW}, \textsc{C-RRW}, and \textsc{C-MBTF}.
Each of these algorithms is stable for any jamming burstiness, unlike the non-adaptive algorithms we considered in Section~\ref{sec:non-adaptive-algorithms-with-jamming}, which include in their codes a bound on jamming burstiness which they can withstand in a stable manner.

First, we estimate the worst-case performance of \textsc{OFC-RRW}, which combines adaptivity with the old-go-first approach, on top of the round-robin-withholding way to use a token.


\begin{lemma}
\label{lem:c-rrw-cycle-plus-packets}

If there are $x$ old packets in the queues, when a phase of algorithm \textsc{OFC-RRW} executed by $n$ stations begins, against a  type $(\rho,\lambda,\beta)$ adversary, then the phase takes at most the following number of rounds:
\[
\frac{x + n }{1-\lambda}
\ .
\] 

\end{lemma}

\begin{proof}
It takes up to $n$ control rounds for the token to pass through all $n$ stations.
It takes $x$ rounds to hear the $x$ packets.
These $x+n$ rounds can be extended to $\frac{x+n}{1-\lambda} $ by the stretching-by-jamming effect.
\end{proof}

Now we give performance bounds for algorithm \textsc{OFC-RRW}.


\begin{theorem}
\label{thm:OFC-RRW-jamming}

If algorithm \textsc{OFC-RRW} is executed by $n$ stations against a jamming adversary of type $(\rho,\lambda,\beta)$ then the number of packets queued in any round is at most
\begin{equation}
\label{eqn:OFC-RRW-jamming-queues}
\frac{2\rho}{1-\rho-\lambda} \cdot n +\beta
\end{equation}
and packet latency  is  at most 
\begin{equation}
\label{eqn:OFC-RRW-jamming-latency}
\frac{2}{1-\rho-\lambda} \cdot n + \frac{\beta(1+\rho-\lambda) }{(1-\lambda)^2}
\ .
\end{equation}
\end{theorem}

\begin{proof}
Let $T_i$ denote an upper bound on the duration of phase~$i$, for $i\ge 1$, where $T_1=\frac{n}{1-\lambda}$, as it consists of $n$ rounds possibly stretched by jamming.
Let $Q_i$ be the number of old packets in the beginning of phase~$i$, for $i\ge 1$.
We use the following two estimates to derive a recurrence for the numbers~$T_i$.
One is
\begin{equation}
\label{eqn:aaa-OFC-RRW-jamming}
Q_{i+1}\le \rho T_i\ , 
\end{equation}
which follows from the definition of old packets and the adversary of type $(\rho,\lambda,b)$.
The other is
\begin{equation}
\label{eqn:bbb-OFC-RRW-jamming}
T_{i+1}\le \frac{Q_{i+1}+n}{1-\lambda} \ ,
\end{equation}
which follows from Lemma~\ref{lem:c-rrw-cycle-plus-packets}.  
Using the abbreviations $c=\frac{n}{1-\lambda}$ and $d=\frac{\rho}{1-\lambda}$, we substitute \eqref{eqn:aaa-OFC-RRW-jamming} into~\eqref{eqn:bbb-OFC-RRW-jamming} to obtain
\[
T_{i+1}
\le
\frac{n}{1-\lambda} + \frac{\rho }{1-\lambda} T_i
\le
c + d T_i\ .
\]
To find an upper bound $T$ on the duration of a phase, we iterate the recurrence $T_{i+1}\le c + d T_i$, which produces
\begin{equation}
\label{eqn:ccc-OFC-RRW-jamming}
 c+d c + d^2 c +\ldots d^i c\le \frac{c}{1-d} =T\ .
\end{equation}
After substituting $c=\frac{n}{1-\lambda}$ and $d=\frac{\rho}{1-\lambda}$ into~\eqref{eqn:ccc-OFC-RRW-jamming}, we obtain the following estimate of the duration of a phase
\begin{equation}
\label{eqn:ddd-OFC-RRW-jamming}
T =
\frac{n}{1-\lambda} \cdot \frac{1}{1-\frac{\rho}{1-\lambda}} 
=
\frac{n}{1-\lambda} \cdot \frac{1-\lambda}{1-\rho-\lambda} 
=
\frac{n}{1-\rho-\lambda}\ .
\end{equation}

A packet spends at most two consecutive phases waiting to be heard.
A phase takes at most~$T$ rounds, where a bound for~$T$ is given in~\eqref{eqn:ddd-OFC-RRW-jamming}.
This bound does not include effects due to burstiness.
To extend it, we can argue as follows.
Let the adversary inject extra $\beta$ packets in a round of a phase.
This extends the duration of the next phase by $\frac{\beta}{1-\lambda}$, because the injected packets will be old then.
Next, the adversary injects extra $\frac{\rho\beta}{1-\lambda}$ additional packets, which are transmitted in the next phase to extend its duration  by $ \frac{\rho\beta}{(1-\lambda)^2}$ rounds, by stretching-by-jamming.

Now we can estimate packet latency as $2T$ incremented by the effects of jamming, to obtain
\[
\frac{2n}{1-\rho-\lambda} + \frac{\beta}{1-\lambda} + \frac{\rho\beta}{(1-\lambda)^2}
=
\frac{2n}{1-\rho-\lambda} + \frac{\beta(1+\rho-\lambda) }{(1-\lambda)^2}
\ ,
\]
which is the bound~\eqref{eqn:OFC-RRW-jamming-latency}.
The number of packets in the stations' queues equals the sum of the numbers of the new and old packets, which is at most $2T\rho$, increased by burstiness to $2T\rho+\beta$, which combined with~\eqref{eqn:ddd-OFC-RRW-jamming} yields the following value
\[
\frac{2n\rho}{1-\rho-\lambda} +\beta
\]
as a bound on the queue size~\eqref{eqn:OFC-RRW-jamming-queues}.
\end{proof}

The bounds of Theorem~\ref{thm:OFC-RRW-jamming} are tight, as can be argued as follows. 
Let the adversary inject at full power.
The first phase takes exactly $n$ clear rounds, which are extended to $\frac{n}{1-\lambda}$ rounds  by jamming at full power.
During this time, the adversary injects $\frac{\rho n}{1-\lambda}$ packets.
So the combined effect of jamming and injecting at full power is injecting with rate $\frac{\rho}{1-\lambda}$. 
The duration of phases keeps increasing such that when one takes $r$ rounds then the next one takes $r(1+\frac{\rho }{1-\lambda})$ rounds.
Eventually, the duration of a phase gets arbitrarily close to 
\[
\frac{n}{1-\lambda} \bigl(1+ \frac{\rho}{1-\lambda} + \bigl(\frac{\rho}{1-\lambda}\bigr)^2+\ldots \bigr) = \frac{n}{1-\lambda} \cdot \frac{1}{1-\frac{\rho}{1-\lambda}} = \frac{n}{1-\rho-\lambda}
\ .
\]
If a phase lasts close to this number of rounds, the adversary injects about $\frac{\rho n}{1-\rho-\lambda}$ packets.  
In one round, the adversary injects $\beta$ packets, which increases the number of packets in the queues to about $\frac{\rho n}{1-\rho-\lambda}+\beta$.
These extra $\beta$ packets then allow the adversary to extend the duration of the next two phases by close to $\frac{\beta}{1-\lambda}+\frac{\rho\beta}{(1-\lambda)^2}=\frac{\beta(1-\lambda+\rho) }{(1-\lambda)^2}$ many rounds.

Next, we estimate the performance of algorithm \textsc{C-RRW}.


\begin{theorem}
\label{thm:C-RRW-jamming}
If algorithm \textsc{C-RRW} is executed by $n$ stations against a jamming adversary of type $(\rho,\lambda,\beta)$ then the number of packets queued in any round is at most
\begin{equation}
\label{eqn:C-RRW-jamming-queues}
\frac{2\rho}{1-\rho-\lambda} \cdot n+\beta
\end{equation}
and packet latency is at most
\begin{equation}
\label{eqn:C-RRW-jamming-latency}
\frac{2(1-\lambda)}{(1-\rho-\lambda)^2}\cdot n+ \frac{\beta(1-\lambda)}{1-\rho-\lambda}
\ .
\end{equation}
\end{theorem}

\begin{proof}
Packets injected behind the token are transmitted in the next phase, which is consistent with the behavior of \textsc{OF-RRW} and so with its bound on the queue size.
Packets injected ahead of the token are transmitted in the phase of injection, which slows down the phase compared to phases of~\textsc{OF-RRW}.
Each such an extra round is spent on a  transmission, because this why a phase is longer, while it is not necessary to have a packet injected in each extra round.
Therefore the upper bound on the number of packets stored in the queues~\eqref{eqn:OFC-RRW-jamming-queues} derived for algorithm \textsc{OFC-RRW} also applies to \textsc{C-RRW}, so it is made equal to~\eqref{eqn:C-RRW-jamming-queues}.

We estimate how injected packets contribute to extending phases by separately considering  packets that are transmitted after a phase of injection and those that are transmitted in a phase of injection.
Packets injected behind the token are transmitted in the next phase, which means that this manner of injecting packets can increase a phase length to be at most as long as the bound of~\eqref{eqn:ddd-OFC-RRW-jamming}, obtained for algorithm \textsc{OFC-RRW}.
Let us denote this bound by $T=\frac{n}{1-\rho-\lambda}$.

Packets injected ahead of the token get transmitted in the phase of injection.
We may assume without loss of generality that when a phase begins the stations store so many old packets that the phase would last $T$ rounds without any additional injections.
If the adversary switches to injecting ahead of the token then combined stretching can extend a phase to at most the following duration:
\begin{equation}
\label{eqn:aaa-C-RRW-jamming}
T \Bigl(1+ \frac{\rho}{1-\lambda}+  \bigl(\frac{\rho}{1-\lambda}\bigr)^2  +\ldots \Bigr)
=  
\frac{T}{1-\frac{\rho}{1-\lambda}}
=
T\cdot\frac{1-\lambda}{1-\rho-\lambda}
\ .
\end{equation}
Substituting $T=\frac{n}{1-\rho-\lambda}$ into~\eqref{eqn:aaa-C-RRW-jamming} produces the following bound on the duration of a phase:
\begin{equation}
\label{eqn:bbb-C-RRW-jamming}
\frac{n}{1-\rho-\lambda}\cdot\frac{1-\lambda}{1-\rho-\lambda}
\le
\frac{n(1-\lambda)}{(1-\rho-\lambda)^2}
\ .
\end{equation}
When such a phase ends, then there are no old packets to be transmitted in the following phase.
Therefore the  lengths of two consecutive phases is at most a sum of~\eqref{eqn:ddd-OFC-RRW-jamming} and~\eqref{eqn:bbb-C-RRW-jamming}, when disregarding the effect of burstiness.
One phase may be further extended by double stretching combined with burstiness by $\frac{\beta(1-\lambda)}{1-\rho-\lambda}$ rounds.
We can conclude with the following bound on packet latency
\begin{equation}
\label{eqn:ccc-C-RRW-jamming}
\frac{n}{1-\rho-\lambda}+\frac{n(1-\lambda)}{(1-\rho-\lambda)^2} + \frac{\beta(1-\lambda)}{1-\rho-\lambda}
\le 
\frac{2n(1-\lambda)}{(1-\rho-\lambda)^2} + \frac{\beta(1-\lambda)}{1-\rho-\lambda} 
\ ,
\end{equation}
which is the bound~\eqref{eqn:C-RRW-jamming-latency}.
\end{proof}

The tightness of the  bounds given in Theorem~\ref{thm:C-RRW-jamming} can be established as follows. 
The queue-size bound is the same as for algorithm \textsc{OFC-RRW}, and the adversary can make the number of packets get close to the bound by making \textsc{C-RRW} behave like \textsc{OFC-RRW} by injecting just behind the token.
In a similar manner, a phase's duration can become close to~\eqref{eqn:ccc-OFC-RRW-jamming}.
Then the adversary switches to injecting at full power  ahead of the token to create one phase of a length close to~\eqref{eqn:bbb-C-RRW-jamming}.
When the second of these two consecutive phases gets extended by burstiness combined with double stretching, the two consecutive phases take time close to the bound in the derivation~\eqref{eqn:ccc-C-RRW-jamming}.

Finally, we estimate the queue sizes and packet latency of algorithm \textsc{C-MBTF} against jamming adversaries.

The relevant terminology we use follows the one developed for executions of algorithm \textsc{MBTF} discussed in Section~\ref{sec:adaptive-algorithms-without-jamming}, in particular, the vocabulary related to kinds of passes.
Similarly, we refer to the stations by their positions on the list of all the stations, numbered $1,2,3,\ldots, n$, with station~$1$ at the front of the list, and station~$n$ at the end.


\begin{lemma}
\label{lem:queues-MBTF-small-jamming}

If algorithm  \textsc{C-MBTF} is executed by $n$ stations against a jamming adversary of type $(\rho,\lambda,\beta)$ such that all the passes have been small since the beginning of the execution,  then the number of packets stored in the queues in a round of a still small pass is at most 
\begin{equation}
\label{eqn:C-MBTF-lemma}
\frac{\rho  n^2}{1-\lambda} + \beta
\ .
\end{equation} 
\end{lemma}

\begin{proof}
A small pass takes $n$ clear rounds, which can be extended to take $\frac{n}{1-\lambda}$ rounds by the stretching-by-jamming argument.
During a time segment of these many rounds the adversary can inject $\frac{\rho n}{1-\lambda}$ packets.

We may assume that a pass under consideration is non-decreasing, as otherwise the last such a pass would witness more packets in the queues.
The quantity $\frac{\rho n}{1-\lambda}$ is an upper bound on the number of stations with packets during a non-decreasing small pass, because if there were more such stations, then each of them would transmit a packet during a pass and more packets would be transmitted than injected.

Each station with packets has at most $n-1$ packets, when it is visited by the token, because no big station has ever been discovered.
It follows that if a small pass is non-decreasing and we disregard burstiness then the number of packets in the queues is at most these many
\[
 \frac{\rho n}{1-\lambda}\cdot(n-1) +  \frac{\rho n}{1-\lambda}
 =
 \frac{\rho  n^2}{1-\lambda} 
\]
when the pass is over.
The adversary can inject extra $\beta$ packets in any round of the passes used to collect these many packets in queues, which justifies the bound~\eqref{eqn:C-MBTF-lemma}.
\end{proof}

A limit that small passes impose on the adversary to build queues is captured by Lemma~\ref{lem:queues-MBTF-small-jamming}.
Big passes may contribute to delaying specific packets by preventing the token to reach the tail of the list of stations over an extended period of rounds.
This may be amplified by the token passing through many stations with empty queues before discovering a big station, while the fraction of void rounds in a pass could be greater than~$\frac{1-\rho}{1-\lambda}$.
During a cascade of such passes,  the adversary may maintain a fraction $\frac{\rho}{1-\lambda}$ of the number of rounds with injections, while the fraction of the number of rounds when messages are heard on the channel is continuously smaller than~$\frac{\rho}{1-\lambda}$, thus contributing to accumulation of packets in the queues.
These insights are employed in the proof of Theorem~\ref{thm:CMBTF-jamming}.


\begin{theorem}
\label{thm:CMBTF-jamming}

If algorithm  \textsc{C-MBTF} is executed by $n$ stations against a jamming adversary of type $(\rho,\lambda,b)$  then the number of packets queued  in any round is at most 
\begin{equation}
\label{eqn:MBTF-packets-jamming}
\frac{\rho(1-\lambda)+\rho^2 }{(1-\lambda)^2} \cdot n^2 + \beta
\end{equation}
and packet latency is at most
\begin{equation}
\label{eqn:MBTF-latency-jamming}
\frac{1+\rho-\lambda -\rho^2-2\rho\lambda}{(1-\lambda)(1-\rho-\lambda)}  \cdot n^2 + \frac{\beta(1-\lambda)}{1-\rho-\lambda}
\ .
\end{equation}
\end{theorem}

\begin{proof}
We initially disregard the effect of burstiness in the course of an execution.
If no big station has been discovered yet then there are at most  $\frac{\rho n^2}{1-\lambda}$ packets in queues, by Lemma~\ref{lem:queues-MBTF-small}.
We explore now how much the queues can increase when big passes occur.
If there are at most $\frac{\rho n}{1-\lambda}$ stations with packets then the sum of lengths of big passes is maximized when the stations with packets are located at the end of the list and when each time the token reaches one of these stations, for the first time since big passes started to occur, then the station is discovered to be big.
The sum of lengths of big passes can be estimated to be at most the following:
\[
\sum_{i=1}^{\frac{\rho n}{1-\lambda}} \bigl(n-\frac{\rho n}{1-\lambda}+i\bigr)
\le
\frac{\rho n^2}{1-\lambda} 
 \ ,
\]
which is the number of clear rounds only.
These big passes can be extended by stretching-by-jamming to last at most $\frac{1}{1-\lambda}\cdot \frac{\rho}{1-\lambda}\cdot n^2$ rounds during which at most $ (\frac{\rho}{1-\lambda})^2\cdot n^2$ new packets are injected. 
The total number of packets at this point is at most
\[
\frac{\rho }{1-\lambda} \cdot n^2 + \frac{\rho^2 }{(1-\lambda)^2} \cdot n^2
=
\frac{\rho(1-\lambda)+\rho^2 }{(1-\lambda)^2} \cdot n^2
\ .
\]
This can be increased to at most~\eqref{eqn:MBTF-packets-jamming} by injecting $\beta$ packets in one round. 

Next we estimate packet latency.
Let us consider a round when some packet $p$ is injected and count the rounds until it is heard.
Burstiness-related effects are disregarded through the initial stages of the analysis, to be accounted for at the end.
We may assume that $p$ gets injected when the token is at the first station and the configuration of packets is as in Lemma~\ref{lem:queues-MBTF-small-jamming}, which is such that at most $\frac{\rho  n^2}{1-\lambda} $ packets are located  in the last $\rho  n$ stations on the list, each holding at most  $n$ packets.
The last station stores the  packet~$p$.
If there are $n-2$ packets preceding $p$ in its queue then~$p$ will be transmitted when the token visits its station  $n-1$ times since $p$'s  injection.
Then the token will need to cover the whole length of the list $n-1$ times to reach~$p$.
Each such a traversal of the whole list makes a small pass, since the last station in the list stores fewer than $n$ packets. 

We estimate how much time may pass before the token visits the $p$'s station when $p$ is at the top of the queue ready to be transmitted, by accounting for the following groups of rounds contributing to $p$'s waiting time: 
\begin{enumerate}
\item[(1)] a delay due to discovering big stations, 

\item[(2)] a delay due to small passes and packets injected during these passes, 

\item[(3)] the effect of burstiness.
\end{enumerate}
We begin with the effect of discovering big stations by identifying a worst-case scenario.
Starting from the injection of~$p$, the adversary may inject packets into the trailing $\frac{\rho n}{1-\lambda}$ stations to make each of them but the last one big.
The discoveries of up to $\frac{\rho  n}{1-\lambda}$ big stations at the end of the list provide delays of up to these many clear rounds:
\[
\sum_{i=1}^{\frac{\rho n}{1-\lambda}} \bigl(n-\frac{\rho n}{1-\lambda}+i\bigr)
\le
\frac{\rho n^2}{1-\lambda} 
\ .
\]
During these big passes, and until the last big station becomes small after moved to the front, the worst case occurs when the duration of waiting time is obtained from the number of clear rounds by combined stretching to at most these many rounds in total:
\begin{equation}
\label{eqn:CMBTF-delay-big}
\frac{\rho n^2}{1-\lambda} \bigl(1+\frac{\rho}{1-\lambda} + \frac{\rho^2}{(1-\lambda)^2} +\ldots\bigr)
=\frac{\rho n^2}{1-\lambda}\cdot \frac{1}{1-\frac{\rho}{1-\lambda}}
=\frac{\rho }{1-\rho-\lambda}\cdot n^2
\ .
\end{equation}

Next, we consider the effect of small passes.
There are $n-1$  small passes before the token reaches~$p$, each requiring $n$ clear rounds, for the total of $n(n-1)$ clear rounds, which can be extended by stretching-by-jamming to at most $\frac{n^2}{1-\lambda}$.
There is also a delay caused by packets injected during small passes.
It is upper bounded by the number of packets injected during small passes, as shown in the proof of Theorem~\ref{thm:MBTF}.
The number of packets injected during small passes is at most $\rho\cdot \frac{n^2}{1-\lambda}$, which makes the total delay incurred by small passes to be at most the following:
\begin{equation}
\label{eqn:CMBTF-delay-small}
\frac{ n^2}{1-\lambda} +\frac{ \rho n^2}{1-\lambda}
= \frac{1+\rho }{1-\lambda} \cdot n^2
\ .
\end{equation}
Burstiness amplified by combined stretching adds at most these many rounds to a big pass:
\begin{equation}
\label{eqn:CMBTF-delay-burstiness}
\beta \bigl(1+\frac{\rho}{1-\lambda} + \frac{\rho^2}{(1-\lambda)^2} +\ldots\bigr)
=
\frac{\beta}{1-\frac{\rho}{1-\lambda}} = \frac{\beta(1-\lambda)}{1-\rho-\lambda}
\ .
\end{equation}
We have identified the three components~\eqref{eqn:CMBTF-delay-big}, \eqref{eqn:CMBTF-delay-small} and~\eqref{eqn:CMBTF-delay-burstiness} of packet delay.
Adding them together gives a total of at most these many rounds:
\begin{eqnarray}
\label{eqn:CMBTF-three-contributions}
\hspace*{-3em}
 \frac{\rho }{1-\rho-\lambda} \cdot n^2 
 +\frac{1+\rho}{1-\lambda}\cdot n^2 
 +\frac{\beta(1-\lambda)}{1-\rho-\lambda}
 &=&
 \frac{\rho(1-\lambda)+(1+\rho)(1-\rho-\lambda)}{(1-\lambda)(1-\rho-\lambda)} \cdot n^2
 +\frac{\beta(1-\lambda)}{1-\rho-\lambda}\\
 \nonumber
 &= &
  \frac{1+\rho-\lambda -\rho^2-2\rho\lambda}{(1-\lambda)(1-\rho-\lambda)} \cdot n^2
 +\frac{\beta(1-\lambda)}{1-\rho-\lambda}
 \ ,
\end{eqnarray}
which is the bound~\eqref{eqn:MBTF-latency-jamming}.
\end{proof}

The bounds given in Theorem~\ref{thm:CMBTF-jamming} are asymptotically tight.
We verify this by giving a specific strategy of an adversary of type $(\rho,\lambda,\beta)$ to make queues grow suitably big and a packet delayed by a suitable number of rounds. 

First, we consider the queues.
The factor $\frac{\rho(1-\lambda)+\rho^2 }{(1-\lambda)^2}$ in the upper bound~\eqref{eqn:MBTF-packets-jamming} is $\Theta(\frac{\rho}{1-\lambda})$ because $1<1+\frac{\rho}{1-\lambda}< 2$.
It is sufficient to show how to construct a configuration with $\Omega(\frac{\rho  }{1-\lambda}\cdot n^2 + \beta)$ queued packets.
Let the adversary begin by building and maintaining queues of $n-1$ packets each in $\frac{\rho}{1-\lambda}\cdot \frac{n}{2}$ selected stations.
This is accomplished during a sequence of $n-1$ small passes of $n$ clear rounds each, which can be extended to $\frac{n}{1-\lambda}$ rounds a pass by the stretching-by-jamming effect.
During each such a pass, the adversary injects $\frac{\rho }{1-\lambda}\cdot n$ packets.
Packets are injected two per station, for as long as there are fewer than $n-1$ packets in each such a station.
Since two packets are injected into a station, while one is transmitted, the number of packets in the stations with packets continues to grow an eventually surpassing the number $\frac{\rho }{1-\lambda}\cdot \frac{n^2}{4}$.
This quantity can be increased to the claimed magnitude by injecting $\beta$ packets in one round.

Next we consider packet latency.
The factor $\frac{1+\rho-\lambda -\rho^2-2\rho\lambda}{1-\lambda}$ in the upper bound~\eqref{eqn:MBTF-latency-jamming} is $\Theta(1)$ because of the following estimates
\[
1
<
\rho+(1+\rho)\bigl(1+\frac{\rho}{1-\lambda}\bigr)
=
\frac{1+\rho-\lambda -\rho^2-2\rho\lambda}{1-\lambda}
<
\frac{1-\lambda +\rho}{1-\lambda} <2
\ ,
\]
where we used $\rho<1-\lambda$.
It is sufficient to show how to create a packet whose delay is  
\[
\Omega\Bigl( \frac{n^2 + \beta(1-\lambda)}{1-\rho-\lambda} \Bigr) = \Omega\biggl(\frac{\frac{n^2}{1-\lambda} + \beta}{1-\frac{\rho}{1-\lambda}}\biggr)
\ .
\]
Let the adversary build queues of $n-1$ packets each in the last  $\frac{\rho}{1-\lambda}\cdot \frac{n}{2}$ stations in the course of a series of small passes.
In each such a pass, the number of packets in a station with packets grows by one, while all the remaining stations have empty queues.
It takes at least $\frac{n^2}{2}$ clear rounds to complete these passes, which can be extended to $\frac{n^2}{2(1-\lambda)}$ by stretching-by-jamming.
A packet $p$ is injected into the last station at the bottom of its queue.
Let the adversary make each of the stations with packets big by inserting one extra packet, starting with the station in the smallest position, but skipping the last station.
After that, the adversary keeps injecting at full power into the station that is last but one.

We consider two cases.
The first case is when $\frac{\rho}{1-\lambda}\le \frac{1}{2}$.
Then the following upper bound holds $\frac{\frac{n^2}{1-\lambda} + \beta}{1-\frac{\rho}{1-\lambda}}\le \frac{2n^2}{1-\lambda} + 2\beta $.
The big passes contribute at least~$\beta$ rounds and the small passes that follow contribute at least $\frac{n^2}{2(1-\lambda)}$ rounds.
The second case is when $\frac{\rho}{1-\lambda} > \frac{1}{2}$.
The big passes generate at least $\sum_{i=1}^{\frac{n}{4}} \bigl(n-\frac{n}{2} +i\bigr) \ge \frac{n^2}{8}$ void rounds in total.
This amount can be extended to at least  $\frac{1}{8}\cdot \frac{n^2}{1-\lambda}$ rounds by stretching-by-jamming.
When the last-but-one station is discovered big, the adversary injects $\beta$ packets into it, to contribute $\beta$ more rounds needed to transmit these packets.
The number of rounds $\frac{1}{8}\cdot \frac{n^2}{1-\lambda}+\beta$ can be extended by double stretching to at least these many rounds
\[
\frac{1}{8}\cdot \frac{n^2+\beta(1-\lambda)}{1-\lambda}\cdot \frac{1-\lambda}{1-\rho-\lambda}
=
\frac{1}{8}\cdot \frac{n^2+\beta(1-\lambda)}{1-\rho-\lambda}
\ .
\]
This quantity grows unbounded if $\frac{\rho}{1-\lambda}$ converges to~$1$.

\section{Conclusion}

\label{sec:conclusion}

We present a comprehensive study  of  distributed deterministic broadcast algorithms in adversarial multiple-access channels, in which an adversary controls packet injection and, optionally, jamming.
The model assumes a fixed set of $n$ stations attached to a shared channel, with the stations equipped with unique names in the interval $[0,n-1]$.
We derive tight asymptotic upper bounds on worst-case queues size and packet latency, which are  expressed in terms of the quantitative constraints defining adversaries and the number  of stations~$n$ using the channel. 

Algorithms are categorized as either adaptive or non-adaptive, channels may either have collision detection mechanism or not, and an adversary may either be able to jam the channel or not.
The case of channels with both jamming and collision detection is omitted from the considerations, in that we do not consider algorithms designed specifically for such an environment.
There are two reasons for this apparent omission.
One is that this model does not allow to distinguish a round with collision form a jammed one, so it is not clear how to make use of collision detection.  
Secondly, such  environments allow to execute any algorithm for  a channel without collision detection and without jamming that avoids collisions, adapted only by the stipulation that jammed rounds, which are detected by a collision-detection mechanism, are to be ignored.

Algorithms \textsc{OF-SRR} and \textsc{SRR}  for channels without jamming but with collision detection have bounds on running times about twice as large as algorithms \textsc{OF-RRW} and \textsc{RRW} for channels without collision detection.
This appears to indicate that utilizing collision detection incurs an additional overhead, which is counter-intuitive, as presence of collision detection is an enhancement of the functionality of a channel when compared to lack of such detection.
The only apparent advantage of collision detection we recognize is to make it possible to have an algorithm of packet latency proportional to burstiness, for injection rates that are $\cO(1/\log n)$.
A similar phenomenon occurs when an adversary is additionally constrained to be able to activate at most one new station in a round, see~\cite{AnantharamuC15}.
In such a communication model, packet latency proportional to burstiness can be attained by deterministic distributed algorithms even for sufficiently small \emph{constant} rates, while the set of stations may be dynamic and names are not needed.

The goals of this work included comparing adaptive algorithms with non-adaptive ones, and investigating the impact of jamming when a jammed round is perceived by stations similarly to a round with collision.
The algorithms we gave appear to demonstrate that these features of algorithms and channels matter, but no impossibility results are given.
Some apparent hypotheses about impossibilities are formulated as conjectures in Section~\ref{sec:introduction}, as part of the review of the results.

This paper considers performance of deterministic distributed broadcast algorithms when measured by their worst-case packet latency.
This performance metric could be considered comparably relevant to real-world applications as  average packet latency. 
Proposing  adversarial models for packet injection to study average packet latency of deterministic distributed broadcast algorithms would be an interesting direction of future work.


\bibliographystyle{abbrv}

\bibliography{mac-latency}

\end{document}